\documentclass[%
 aip,
 amsmath,amssymb,
 reprint,%
]{revtex4-2}

\usepackage{graphicx}
\usepackage{dcolumn}
\usepackage{bm}

\usepackage[utf8]{inputenc}
\usepackage[T1]{fontenc}

\usepackage[%
	breaklinks=true,
	colorlinks=true,
	linkcolor=blue,
	urlcolor=blue,
	citecolor=blue
]{hyperref}
\usepackage{algorithm}
\usepackage{algpseudocode}
\usepackage{algorithmicx}
\DeclareMathOperator\arctanh{arctanh}
\usepackage{mathptmx}
\usepackage{etoolbox}

\makeatletter
\def\@email#1#2{%
 \endgroup
 \patchcmd{\titleblock@produce}
  {\frontmatter@RRAPformat}
  {\frontmatter@RRAPformat{\produce@RRAP{*#1\href{mailto:#2}{#2}}}\frontmatter@RRAPformat}
  {}{}
}%
\makeatother
\begin{document}

\preprint{AIP/123-QED}

\title[Preprint]{Simulation of charged nanotubes self-assembly during evaporation of a sessile droplet on a substrate}
\author{Konstantin~S.~Kolegov} \email{konstantin.kolegov@asu-edu.ru}
\affiliation{Laboratory of Mathematical Modeling, Astrakhan Tatishchev State University, Russia.}
\author{Irina V. Vodolazskaya}
\affiliation{Laboratory of Mathematical Modeling, Astrakhan Tatishchev State University, Russia.}
\author{Andrei V. Eserkepov}
\affiliation{Laboratory of Mathematical Modeling, Astrakhan Tatishchev State University, Russia.}
\author{Ludia T. Khusainova}
\affiliation{Laboratory of Mathematical Modeling, Astrakhan Tatishchev State University, Russia.}

\date{\today}

\begin{abstract}
The ability to control the morphology of the nanotube deposit formed during the evaporation of a sessile droplet on a substrate is of theoretical and practical interest. Such deposits are required for various applications including nanotechnology, medicine, biotechnology, and optronics. In the experiment of Zhao \textit{et al.} [\href{https://www.doi.org/10.1016/j.jcis.2014.10.050}{J. Colloid Interface Sci. 440, 68 (2015)}], an annular deposit was formed near the contact line. The deposition geometry is caused by the coffee ring effect. This deposit is unusual in its morphology. It changes gradually in space from a disordered structure in the inner part of the ring to an aligned structure of nanotubes close to the periphery. To understand the mechanisms that lead to this, we have developed a mathematical model that takes into account the effects of advection, diffusion, and electrostatic interactions on particle transport. Results of numerical calculations have confirmed that all these factors together have an influence on the formation of such a variable morphology. Qualitative agreement with the experiment is shown for some values of the model parameters.
\end{abstract}

\maketitle

\section{Introduction}
Evaporation-induced self-assembly has promising applications in micro- and optoelectronics. For example, transparent electrically conductive films can be created from carbon nanotubes using inkjet printing~\cite{Shimoni2014}. This method is based on the coffee ring effect. In an experimental study~\cite{Inanlu2022} using this method, the possibility of creating a strain gauge was described. Different concentrations of surfactants in a solution containing carbon nanotubes were used and the concentration at which a wide annular deposit forms was determined. To increase the density of these deposits and improve electrical properties, droplets are applied alternately at the same location after the previous droplet has dried. This results in the formation of a dense multilayer ring deposit~\cite{Inanlu2022}. In another experiment~\cite{Howard2023}, the effect of surfactant concentration and substrate material on the geometry of carbon nanotube deposits was studied. Different forms of deposition were obtained, including dot-like deposits, uniform deposits, coffee ring patterns, and combined patterns (coffee ring with a central dot-like deposit). Inkjet printing enables the deposition of relatively long, narrow layers of carbon nanotubes~\cite{Denneulin2011,Naderi2024}. The authors~\cite{Denneulin2011} studied the morphology of deposits formed near the contact line and measured their resistance. Thin-film transistors can be created based on these layers~\cite{Naderi2024}.

In the experiment with a capillary bridge formed by a confined liquid layer between the substrate and the cover glass~\cite{LiH2014}, carbon nanotubes were deposited during the retraction of the contact line. The contact line moved in a stick-slip mode, leading to the formation of stripe-like deposits on the substrate. At the same time, there was periodic uneven detachment of the contact lines after the stage of pinning and spreading of its kink, with some particles being captured by the contact line during compression and others being transported by capillary flow caused by evaporation toward this boundary. Several parameters were studied: the height of the capillary bridge, the concentration of nanotubes and surfactants~\cite{LiH2014}. For coating and electronic component applications, it is sometimes useful to form a network of carbon nanotubes (uniformly distributed nanotubes)~\cite{Small2006}.The authors studied the effect of hydrophobic substrates and pre-centrifugation of the solution on the formation of deposits in the area of the dried droplet.

The evaporative self-assembly of halloysite nanotubes was studied in a paper~\cite{Zhao2015}. In their experiments, the authors observed the effect of coffee rings at the place where dried droplets were on the substrate. Key parameters such as the charge of the nanotubes, the temperature of the liquid, the aspect ratio of the nanotubes and the concentration of the solution were taken into account. At certain values of these parameters, a transition from chaos to ordering in the morphology of the annular deposit occurred. Near the contact line, the orientation of the nanotubes was parallel to this boundary (nematic phase), and closer to the center there was a disordered deposition zone (isotropic phase). There was also an intermediate zone (transition phase). In another experiment with carbon nanotubes, the authors~\cite{Goh2019} studied the effect of particle concentration and substrate temperature on deposit morphology. Theoretically~\cite{Onsager1949} and experimentally~\cite{Zhang20102107}, it was previously shown that the transition from an isotropic to a nematic phase occurs for charged rod-like particles, including carbon nanotubes, when the critical concentration of the solution is exceeded. The aim of this study is to numerically investigate the transition from a disordered to an aligned structure in order to gain a better understanding of the mechanisms that influence the morphology of nanotube deposits. The morphology of halloysite nanotube deposits is important, for example, in the development of membranes for dye/salt separation~\cite{Qin2016}. Halloysite clay tubes have been used as natural nanocontainers for loading and sustained release of chemical agents~\cite{Lvov2015}.

\newpage

\section{Methods}

\subsection{Physical problem formulation}

Consider a water droplet with suspended halloysite nanotubes evaporating on a solid horizontal impermeable substrate. The physical and geometric parameters of the problem are presented in Table~\ref{tab:ParametersInNanotubesSelfAssembly}. The initial height of the drop $h_0$ is determined approximately from an experimental image (Fig. 2(b) from Ref.~\cite{Zhao2015}). Since the ratio between the height of the droplet and the radius of its base is small, $h_0/ R\approx 0.3$, the droplet can be treated as a thin layer. The mass of a nanotube, $m$, was roughly estimated by multiplying its density by the volume of a cylinder-shaped tube. Throughout the process, the contact line remained pinned, as observed in Ref.~\cite{Zhao2015}, where the radius of the droplet base remained constant. In this case, non-uniform evaporation along the free surface of the droplet leads to capillary flow directed from its center to the periphery. This flow transfers the particles closer to the contact line, where they are gradually deposited onto the substrate (the coffee ring effect~\cite{Deegan2000}). No Marangoni flow was detected in experiments using sessile water droplets with nearly identical geometry under comparable room conditions~\cite{HuLarson2006,Hamamoto2011,Zhao2015}.

However, in this case, the situation is complicated by additional factors. First, the particles are charged, which affects their dynamics. Second, they have an elongated shape. This particle shape affects their transport by flow and diffusion. In the experiment~\cite{Zhao2015}, the nanotubes of halloysite clay after functionalization had a large surface charge and were repelled, i.e., their surface charge was not completely shielded in water.

\begin{table}[h]
	\caption{Physical and geometric parameters.}
	\centering
	\begin{tabular}{|p{0.13\linewidth}|p{0.45\linewidth}|p{0.2\linewidth}|p{0.12\linewidth}|}
		\hline
		Symbol& Parameter& Value& Unit of measure\\
		\hline
		$R$& Radius of the droplet base~\cite{Zhao2015} & 1.25 & mm\\
		$\theta_0\approx 2h_0/ R$& Initial contact angle & 0.6 & rad\\
		$l$& Nanotube length~\cite{Zhao2015} & 1.5 & µm\\
		$\epsilon$& Nanotube aspect ratio~\cite{Zhao2015} & 30 & \\
		$h_0$& Initial droplet height & 0.37 & mm\\
		$m$& Nanotube mass & $6.5\times 10^{-18}$ & kg\\
		$T$& Liquid temperature~\cite{Zhao2015} & 338 & K\\
		$\eta_0$& Dynamic viscosity of water~\footnote{At a temperature of $T$.} & $0.44\times 10^{-3}$ & Pa~s\\
        $\rho$& Water density~\footnotemark[1] & $980$ & kg/m$^3$\\
        $k_f$& Viscous friction coefficient (Appendix~\ref{appendix:frictionCoefficient}) & $5\times 10^{-10}$ & kg/s\\
		\hline
	\end{tabular}
	\label{tab:ParametersInNanotubesSelfAssembly}
\end{table}

\subsection{Mathematical model}

The process is described using a two-dimensional (2D) formulation in the
$xy$-plane, parallel to the substrate. Within the scope of this model, we will discuss the formation of a monolayer of nanotubes. However, it should be noted that multilayer deposits can also form in reality~\cite{Park2024}. The effect of the transition from an isotropic to a nematic phase during the evaporation of a droplet can be explained within the scope of such simple geometry (see the sketch in Fig. 3 of Ref.~\cite{Zhao2015}). The review~\cite{Wang2022} on phase states of 2D elongated particle structures provides arguments in favor of the importance of studying such systems.

Within the scope of our model, we assume that nanotubes are characterized by a quasi-zero thickness. When checking for the absence of particle overlap (collisions), we assume that the nanotubes are infinitely thin in the horizontal plane. However, the particle thickness is taken into account in order to control the locations where the thickness of the liquid layer and the nanotube diameter are similar (vertical plane). This allows us to track the moment when a nanotube touches the substrate. The approximation ``zero-width objects'' is often used for cases where $d\ll l$ (see, for example, the Ref.~\cite{Lebovka2019}).

In this paper, a hybrid model is proposed for splitting by physical processes. Individual related processes can be approximated using different methods. Particle diffusion is modeled here by the Monte Carlo method.  The direction of particle displacement and rotation at each time step is randomly chosen. The transfer of particles by flow, including their rotation, is described separately. Due to the low concentration and small size of the nanotubes, it is assumed that the particles have no effect on the fluid flow (see Sec. 2 in Supplemental Material A). When describing the electrostatic interaction between nanotubes within a single time step, we abstract from other transfer processes occurring simultaneously, but also take into account viscous friction. In the model, the different processes within a time step are executed sequentially. But at a sufficiently small time step ($\Delta t \to 0$), this model allows us to mimic the parallel processes in a real physical system.

Let us consider the charge of nanotubes to be an effective value and a variable parameter. Here, we investigate the effect of this parameter on the ordering of the nanotubes near the droplet edge.

\subsubsection{Droplet shape and fixing radius}

\begin{figure}
	\includegraphics[width=0.9\linewidth]{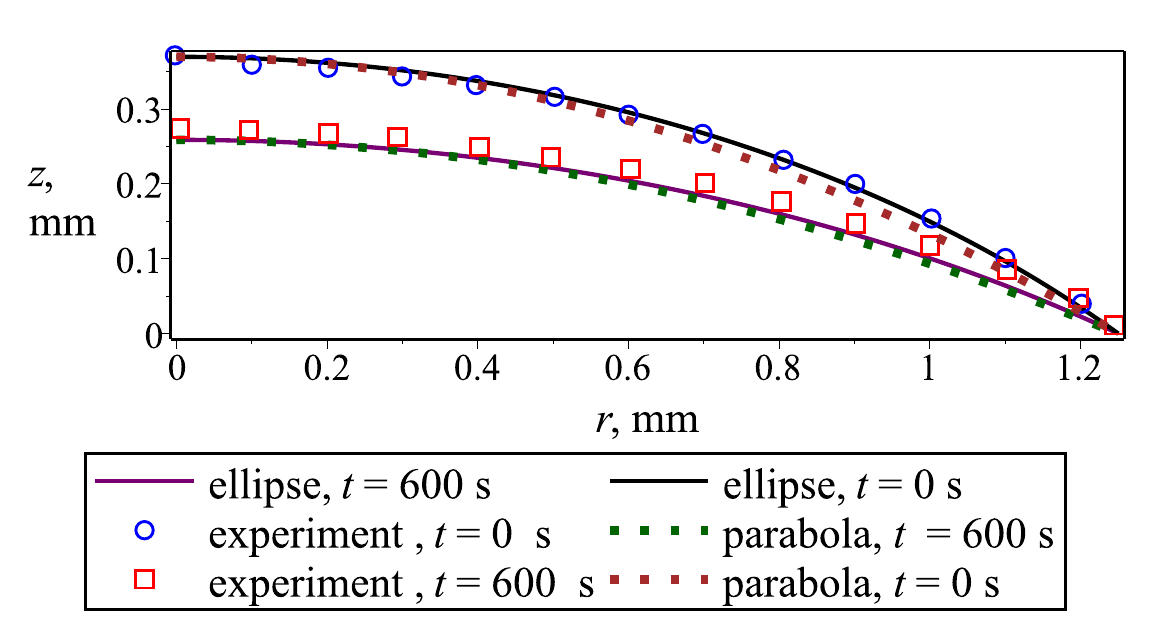}
	\caption{\label{fig:dropShape_experVsFit}Approximation of the droplet shape by parabola and ellipse (experimental data~\cite{Zhao2015}).}
\end{figure}

Choose a formula to approximate the free surface of the droplet. Let's check the parabolic approximation, formula~(1) in Ref.~\cite{Kolegov2019}, and the ellipsoidal approximation, formula~(1) in Ref.~\cite{Gatapova2014}. We have determined the droplet shape using an experimental image (Fig. 2( b) in Ref.~\cite{Zhao2015}) for the moment in time $t=$ 0~s and $t=$ 600~s. The radius of the droplet base (its value), $R$, is presented in Table~\ref{tab:ParametersInNanotubesSelfAssembly}. A comparison of the results is shown in Fig.~\ref{fig:dropShape_experVsFit}. Both approximations provide similar values and are in good agreement with the experiment~\cite{Zhao2015}. However, the parabolic form is computationally more efficient~\cite{Kolegov2019}. Considering the parabolic shape, the expression for a fixing radius, $R_f$,  was previously obtained~\cite{Kolegov2019}. It changes over time during evaporation of the droplet (see the sketch in Fig. 2 of Ref.~\cite{Kolegov2019}),
\begin{equation}\label{eq:FixingRadiusNanotube}
	R_f (t) = \sqrt{R^2 - \frac{2 d R}{\theta(t)}},
\end{equation}
where $\theta$ is the contact angle, $\theta(t) = \theta_0 (1-t/t_\mathrm{max})$,  $t_\mathrm{max}$ is the droplet evaporation time, and $d = l/ \epsilon$ is the particle (cylinder) diameter. On this boundary, $R_f$, where the thickness of the liquid layer is commensurate with that of the nanotube, the particle is affected by the surface tension of the liquid. Thus, the nanotube is pressed against the substrate by a two-phase boundary (the ``liquid--air'' interface). This model assumes the nanotube in contact with the substrate remains fixed (stationary). It is worth noting that the formula~\eqref{eq:FixingRadiusNanotube} is limited by the condition $t \leqslant
t_\mathrm{cri} = t_\mathrm{max} (1 - 2 d /(\theta_0 \, R))$. Otherwise we get a negative number under the root. In our case, the critical time is extremely close to the time of full evaporation, $t_\mathrm{cri} \approx 0.9999\, t_\mathrm{max}$. In addition, if the local height of a droplet is denoted as a function $h(r,t)$, then the nature of the boundary $R_f$ can be written mathematically as $h(R_f(t),t) = d$, where $r$ is the radial coordinate.

\subsubsection{Particle diffusion}

When modeling nanotube diffusion in liquid, we assume a dilute solution. Diffusion can be divided into two components: translational, with a diffusion coefficient $D_t = k_B T/ f_t$ (m$^2$/s), and rotational, with a diffusion coefficient $D_r = k_B T/ f_r$ (1/s)~\cite{Li2004}. In turn, translational diffusion is also conveniently divided into two components: along, $D_{\parallel}= k_B T/ f_{\parallel}$, and perpendicular, $D_{\perp}= k_B T/ f_{\perp}$, to the nanotube axis (Boltzmann constant $k_B\approx$ $1.38\times 10^{-23}$ J/K)~\cite{Li2004}. Now let's write the expressions for the Stokes friction coefficients~\cite{Li2004}:
$$f_t = \frac{8 \pi \eta_0 l}{3 \ln(\epsilon) + 2 \gamma_{\parallel} + \gamma_{\perp}},$$
$$f_{\parallel} = \frac{2 \pi \eta_0 l}{\ln(\epsilon) + \gamma_{\parallel}},$$
$$f_{\perp} = \frac{4 \pi \eta_0 l}{\ln(\epsilon) + \gamma_{\perp}},$$
$$f_{r} = \frac{\pi \eta_0 l^3}{3(\ln(\epsilon) + \gamma_r)}.$$
Correction factor values ($\gamma_{\perp}$,  $\gamma_{\parallel}$, and $\gamma_r$) are given in the paper~\cite{Li2004} for the case $\epsilon \to \infty$. For our simulations, we used the following empirical dependencies that are suitable for this case, $2 \leqslant \epsilon  \leqslant 30$~\cite{Tirado1984,Lowen1994}:
$$\gamma_{\perp} = 0.839+0.185/\epsilon + 0.233/\epsilon^2,$$
$$\gamma_{\parallel} = -0.207+0.98/\epsilon-0.133/\epsilon^2,$$
$$\gamma_r = -0.662+0.917/\epsilon-0.05/\epsilon^2.$$
The displacement distance of the nanotube along its axis is calculated as $\Delta r_{\parallel} = \sqrt{2 D_{\parallel} \Delta t}$~\cite{Li2004}. To determine the perpendicular displacement, one can use the formula $\Delta r_{\perp} = \Delta r_{\parallel} \sqrt{f_{\parallel} / f_{\perp}}$~\cite{Lebovka2019}. The rotation angle of the nanotube is expressed as  $\Delta \alpha = \sqrt{2 D_r \Delta t}$~\cite{Li2004}. The position of the nanotube is determined by its center of mass ($x_C$, $y_C$), and the angle $\alpha$, which the nanotube forms with the $x$-axis (counting from the axis in the positive direction). The coordinates of the nanotube edges are recalculated using the formulas $x_{1,2}=x_C \pm \Delta x$ and $y_{1,2}=y_C \pm \Delta y$, where $\Delta x = 0.5 l \cos (\alpha)$ and $\Delta y = 0.5 l \sin (\alpha)$. The new coordinates of the nanotube after diffusion displacement and rotation are defined as
\begin{equation}\label{eq:DiffusionDisplacementAndRotation}
	x_C = x_C + \Delta x_C, \, y_C = y_C + \Delta y_C, \alpha = \alpha + d_r \Delta \alpha,
\end{equation}
where $\Delta x_C = d_\parallel \Delta r_{\parallel} \cos (\alpha) - d_\perp \Delta r_{\perp} \sin (\alpha)$ and $\Delta y_C = d_\parallel \Delta r_{\parallel} \sin (\alpha) + d_\perp \Delta r_{\perp} \cos (\alpha)$. Here, the parameters $d_r$, $d_\parallel$ and $d_\perp$ define the direction of rotation/displacement. Their values at each time step are set randomly, taking the value of $\pm 1$ with equal probability.

\subsubsection{Effect of the fluid flow}
Let us consider separately the possible types of motion of the nanotube under the influence of the fluid flow: (I) translational motion, (II) free rotational motion relative to the center of mass of the particle, and (III) rotational motion relative to a point of contact with another particle or the fixing radius, $R_f$. The radial flow velocity averaged over the thickness of the liquid layer will also be described, as was done earlier~\cite{Kolegov2019}:
\begin{equation}\label{eq:velocityAnalytical}
	v_r= \frac{R}{4\tilde r (t_\mathrm{max}-t)}\left[  \frac{1}{\sqrt{1-\tilde r^2}} - \left( 1-\tilde r^2 \right)
	\right],
\end{equation}
where $\tilde r = r/R$,  $r=\sqrt{x^2 + y^2}$.

At each $i$-th time step, the particle undergoes translational motion (I) with the flow velocity $\vec{v}$ at the point where the center (point $\mathrm C$) of the nanotube mass is located. The velocity of the nanotube is recorded as a vector $\left. \vec{v}^\mathrm{rod} = \vec{v}\left(\mathrm C \right)\right|_{i}$. Let's write the velocity projections on the axis of the Cartesian coordinate system, $v_x = v_r\, x_\mathrm C/ r$ and $v_y = v_r\, y_\mathrm C/ r$. Over a period of time $\Delta t$ the nanotube moves by $\left. v^\mathrm{rod}_{x} \right|_{i}\times \Delta t $ along the $OX$-axis and by $\left. v^\mathrm{rod}_{y}\right|_{i} \times \Delta t $ along the $OY$-axis. Therefore, we use the following formulas in the calculation
\begin{equation}\label{eq:AdvectionDisplacement}
	\left. x_\mathrm C \right|_{i+1} = \left. x_\mathrm C \right|_{i} + \Delta t \,\left. v^\mathrm{rod}_{x}\right|_{i}, \; \left. y_\mathrm C \right|_{i+1} = \left.y_\mathrm C \right|_{i} + \Delta t \, \left. v^\mathrm{rod}_{y} \right|_{i}.
\end{equation}

The authors~\cite{Goh2019} hypothesized that liquid flow induces nanotube rotation during convective transport toward the droplet periphery. According to this hypothesis, nanotubes should gradually align along the flow direction. Our numerical estimate has indicated that the torque is negligible, as the length of the nanotube is relatively short. Therefore, free rotational motion around the center of mass (II) can be ignored (see Appendix~\ref{appendix:FlowRotation}).

When a nanotube edge contacts either another particle or the fixing radius $R_f$, its subsequent motion is constrained to rotation about the contact point $A_{1}$ (III). The rotation angle of the nanotube during time $\Delta t$, around its edge (point $A_1$), on the current time step, is determined by the formula [see Appendix A, formula~\eqref{eqAeight}]
\begin{multline}
	\label{eq:RotationAngleByFlowRelativeToEnd}
	\left. \Delta \beta \right|_{i} \approx \left. \omega_{z} \right|_{i} \Delta t \approx \\
	\left. \frac{3 \Delta t}{l^{2}} \left(  \left(x_{C} - x_{A_{1}}   \right)v_{y}(x_{C},\, y_{C})
	- \left(y_{C} - y_{A_{1}}   \right)v_{x}(x_{C},\, y_{C})      \right) \right|_{i} .
\end{multline}
The rotation is performed counterclockwise if $\Delta \beta > 0$, and clockwise otherwise.

If a nanotube touches another particle, not with its edge, but at some point $N$, then the rotation of the nanotube in time $\Delta t$ around the point of contact at the calculated time step can be determined as [see Appendix A, formula~\eqref{eqAeleven}]
\begin{multline}
	\label{alphaflownotend}
	\left. \Delta \beta \right|_{i}\approx \frac{\Delta t}{l \left(\left| CN \right|^{2} + l^{2}/ 12 \right)}\times \\ \times\left(\left|A_{1}\,N  \right| \times \left( \left(x_{Q} - x_{N}   \right)v_{y}\left(x_{Q},\, y_{Q}\right)
	- \left(y_{Q} - y_{N}   \right)v_{x}(x_{Q},\, y_{Q})\right)\right.+ \\
	+\left.\left. \left|A_2\,N  \right| \times \left(\left(x_{P} - x_{N}   \right)v_{y}(x_{P},\, y_{P})
	- \left(y_{P} - y_{N}   \right)v_{x}(x_{P},\, y_{P})\right)\right)\right|_{i},
\end{multline}
where points $P$ and $Q$ are the midpoints of the segments into which the nanotube is divided by point $N$. Formula~\eqref{eq:RotationAngleByFlowRelativeToEnd} is obtained from Formula~\eqref{alphaflownotend} in the limit when $N\to A_1$.

\subsubsection{Electrostatic interaction and viscous friction}

Within the scope of this proposed model, the motion of a particle in the flow is superimposed onto the motion of the particle relative to the liquid under the influence of both electrostatic and viscous forces. Consider separately the possible types of motion of the nanotube under the influence of electrostatic forces: (EI) translational motion, (EII) free rotational motion relative to the center of mass of the particle, and (EIII) rotational motion relative to a point of contact with another particle or the fixing radius, $R_f$.

Suppose that at each small time step, the nanotube under consideration undergoes translational motion (EI) with a constant velocity $\vec{u}^\mathrm{rod}$ (steady motion is the result of the drag force balancing the electrostatic force):
$$\vec{F} - k_{f} \vec{u}^\mathrm{rod} = 0, $$
\begin{equation}\label{eq:particleVelocityOnFrictionAndElectrostatic}
	\Rightarrow \quad \left. u^\mathrm{rod}_{x} \right|_{i} = \left. \left( \sum\limits_j F_x^j \right) / k_{f} \right|_{i}, \, \left. u^\mathrm{rod}_{y} \right|_{i} = \left. \left( \sum\limits_j F_y^j \right)  / k_{f} \right|_{i}.
\end{equation}
Here, $\vec{F}^{j}$ is the electrostatic force acting on the current nanotube from the $j$-th particle. For each particle, we calculate the resultant force by summing interactions with all adjacent nanotubes. The viscous friction force is considered to be proportional to the first power of the particle velocity relative to the medium. The viscous friction coefficient  $k_f$ is described in detail in Appendix~\ref{appendix:frictionCoefficient}. To simplify calculations, we assume that the charge of a nanotube is concentrated at its ends (two identical point charges, $q$). The exact formula for calculating the electrostatic force in the case when the charge is evenly distributed over the nanotube can be found in Supplemental Material~A (Sec. 1). However, implementing this formulation presents significant computational challenges in practice.
\begin{figure}
	\centering
	\includegraphics[width=0.6\columnwidth]{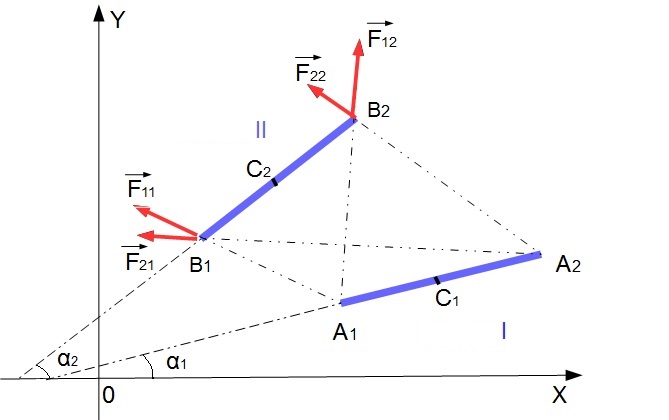}
	\caption{Two charged identical particles on a plane form a system of four identical point charges at the ends of nanotubes. The forces acting on the first particle charges from the second particle charges are shown.}
	\label{fig:ParticleElectrostaticInteraction}
\end{figure}

Figure~\ref{fig:ParticleElectrostaticInteraction} shows a special case of electrostatic interaction between two particles.

The force exerted by nanotube $I$ ($A_{1}\, A_{2}$) on nanotube $II$ ($B_{1}\, B_{2}$) is equal to the sum of four forces, each of which is determined by Coulomb's law
\begin{equation}\label{eq:SumOfForces}
	\vec{F}= \vec{F}_{11} +  \vec{F}_{21} + \vec{F}_{22} + \vec{F}_{12}.	
\end{equation}
For example, $\vec{F}_{11} = k q^2 \frac{\overrightarrow{A_{1} B_{1}}}{|A_{1} B_{1}|^{3}}$, where $k$ is the constant in Coulomb's law. The rest of the forces are recorded in the same way. During the time interval $\Delta t$, the particle moves by $\left. u^\mathrm{rod}_{x} \right|_{i}\times \Delta t $ along the $OX$ axis and by $\left. u^\mathrm{rod}_{y}\right|_{i} \times \Delta t $ along the $OY$ axis.

Consider the free rotational motion relative to the center of mass of the nanotube (EII). Let us write the moment of forces $\vec{M}$, acting on particle $II$ ($B_{1}\, B_{2}$) from the neighboring particle $I$ ($A_{1}\, A_{2}$), relative to the center of the nanotube corresponding to point $C_{2}$ (see Fig.~\ref{fig:ParticleElectrostaticInteraction}),
\begin{multline*}
	\hfill \vec{M} =  \left[\overrightarrow{C_{2}B_{1}},\,\vec{F}_{1\,1} +\vec{F}_{2\,1} \right] + \left[\overrightarrow{C_{2}B_{2}}, \, \vec{F}_{2\,2} + \vec{F}_{1\,2}\right] = \hfill \\ 
	\vec{n}_{z} \left(\left(x_{B1}- x_{C2}  \right)\left(F_{11y}+ F_{21y}  \right)-\left(y_{B1}- y_{C2}  \right)\left(F_{11x}+ F_{21x}  \right)\right).
\end{multline*}

If particle $II$ interacts with several nanotubes, then the moments of forces are summed up. In addition, the torque of the fluid resistance relative to the center of mass of the nanotube is also taken into account, which may be calculated using the formula $M_\mathrm{vis} \approx l^2 k_f \omega_z / 4$, where $\omega_z$~ is the angular rotational velocity of the nanotube (see Appendix~\ref{appendix:frictionCoefficient}).

Let us assume that at each small time step $i$ the moment of resistance balances the moment of electrostatic force, $\sum_j \vec{M}^j = \vec{M}_\mathrm{vis}|_i$. Then, the particle $B_{1}\, B_{2}$ rotates around the center of mass at a constant angular velocity at each time step:
\begin{equation}
	\label{eq:AngularVelocityByViscosityAndElectrostatics}
	\left. \omega_{z} \right|_i = \left. \frac{\left[\overrightarrow{C_2 B_1},\, \sum\limits_j \left(\vec{F}^j_{11} +\vec{F}^j_{21}\right) \right]_z + \left[\overrightarrow{C_2 B_2}, \,\sum\limits_j \left(\vec{F}^j_{22} + \vec{F}^j_{12} \right)\right]_{z} }{l^2 k_f/4}\right|_i.
\end{equation}

During the time $\Delta t$ the nanotube rotates by an angle
$$
	\left. \Delta \alpha \right|_i = \Delta t \omega_{z}|_i,
$$
where $\omega_{z}|_i$ is calculated using the formula~\eqref{eq:AngularVelocityByViscosityAndElectrostatics}. The rotation is performed counterclockwise at $\Delta \alpha > 0$ and clockwise otherwise.

Now consider the rotational motion relative to the point of contact with another particle or the fixing radius, $R_f$ (EIII). If the nanotube $B_{1}\, B_{2}$ at some point in time contacts with its edge $B_{2}$ the fixing radius or another particle, then its further motion is only rotational motion around the point of contact. In this case, the angular velocity of rotation at each time step is determined by the formula
\begin{equation}\label{eq:AngularVelocityByViscosityAndElectrostaticsAgainstItsEnd}
	\left. \omega_z \right|_{i}= \left.\frac{\left.\left[\overrightarrow{B_2 B_1},\, \sum\limits_j \left(\vec{F}^j_{11} +\vec{F}^j_{21}\right) \right]\right|_{z}}{l^{2} k_f}\right|_{i}.
\end{equation}

If the nanotube $B_{1}\, B_{2}$ touches another particle not with its edge, but with some point $N$, then the angular velocity of rotation around point $N$ at the current time step is defined as
\begin{equation}\label{eq:AngularVelocityByViscosityAndElectrostaticsAgainstArbitraryPoint}
	\left. \omega_z \right|_i= \left.\frac{\left.\left[\overrightarrow{N B_1},\, \sum\limits_j \left(\vec{F}^j_{11} +\vec{F}^j_{21}\right) \right]\right|_z + \left.\left[\overrightarrow{N B_2},\, \sum\limits_j \left(\vec{F}^j_{12} +\vec{F}^j_{22}\right) \right]\right|_z }{l^2 k_f/2}\right|_i.
\end{equation}

Therefore, the rotation angle of the nanotube during time $\Delta t$ around the point of contact at the current time step is defined as
$ \left. \Delta \beta \right|_i =  \Delta t \omega_z|_i$.

\subsection{Calculation algorithm}

As variable parameters in this paper, we consider next of them: the number of particles, $N_p$, the droplet evaporation time, $t_\mathrm{max}$, and the charge of nanotubes (parameter $k q^2$, where $k$ is the constant in Coulomb's Law, and $q$ is the charge on the nanotube). These parameters serve as user-defined inputs to the simulation program. With $N_p$ we can vary the initial concentration of the solution. The charge of nanotubes determines the strength of their electrostatic interaction. The time $t_\mathrm{max}$ affects the capillary flow velocity and the advective mass transport. The disadvantage of the analytical formula~\eqref{eq:velocityAnalytical} for the flow velocity is the singularity at $r\to R$ and $t\to t_\mathrm{max}$. During the creation of the program, we take this feature into account and use a modified formula
 \begin{equation}\label{eq:velocityAnalyticalModified}
 	v_r(r,t) =
 	\begin{cases}
 		v_r(r,t), &(r \leqslant r_\mathrm{cut})\wedge (t \leqslant t_\mathrm{cut});\\
 		v_r(r_\mathrm{cut},t_\mathrm{cut}), &(r > r_\mathrm{cut})\wedge (t >t_\mathrm{cut});\\
 		v_r(r_\mathrm{cut},t), &(r > r_\mathrm{cut})\wedge \neg(t > t_\mathrm{cut});\\
 		v_r(r,t_\mathrm{cut}), &\neg(r > r_\mathrm{cut})\wedge (t > t_\mathrm{cut}).\\
 	\end{cases}
 \end{equation}
In Formula~\eqref{eq:velocityAnalyticalModified}, we use the values of the parameters $r_\mathrm{cut}$ and $t_\mathrm{cut}$, so that the maximum value of flow velocity $v_\mathrm{max}$ is of the same order of magnitude as the value obtained from the experiment with a water droplet located on a substrate~\cite{Hamamoto2011} ($R \approx$ 0.45~mm, $t_\mathrm{max}\approx$ 110~s). In Ref.~\cite{Hamamoto2011}, the measurements showed $v_\mathrm{max} \approx$ 60~$\mu$m/s, while Formula~\eqref{eq:velocityAnalyticalModified} outputs the value $v_\mathrm{max} \approx$ 97.5~$\mu$m/s at the corresponding values of $R$ and $t_\mathrm{max}$ for the parameters $r_\mathrm{cut}= 0.85 R$ and $t_\mathrm{cut} = 0.98 t_\mathrm{max}$ used here in the calculations. The effect of the parameters $r_\mathrm{cut}$ and $t_\mathrm{cut}$ on the result is discussed in Supplemental Material~A (see comments to Fig. 6 in Sec. 3).

The calculation of the total area of the droplet takes too long due to the large number of particles, $N_p$ (several months). Perhaps using parallel algorithms and computations on a supercomputer would help solve this problem. Given the axial symmetry of the problem, we can limit ourselves to a specific sector of the circle with periodic boundary conditions. In the calculations, we use the sector with the angle of $\delta = \pi/ 128$ (the effect of $\delta$ on the simulation results is discussed in Supplemental Material~A, Sec.~3). To simplify the implementation of the calculation of the electrostatic interaction force between particles near the boundaries of the sector, we use two adjacent sectors with nanotubes duplicating the main particles in the central sector.

The smaller the value of the time step $\Delta t$ used in the calculation, the more accurate the results obtained. However, at the same time, the longer the execution time of the program will be. Taking into account both factors (accuracy of calculations and execution time), we chose the optimum value of $\Delta t = 10^{-4}$ s for us. With this value, our calculations take less than two weeks  using Intel(R) Xeon(R) CPU E5-2690 v3 2.6 GHz. We have compared some results for special cases with similar results obtained using $\Delta t = 10^{-5}$ s and have not noticed any qualitative differences (see Supplemental Material~A, Sec. ~3). At $\Delta t = 10^{-5}$ s, the calculation takes about 10 times longer. The calculation time is also influenced by other parameters, such as $N_p$, $t_\mathrm{max}$ and $k q^2$. The pseudocode of the algorithm is presented in listing~\ref{alg:NanotubeDynamics}.

\begin{algorithm}[H]
	\caption{Nanotube dynamics calculation} \label{alg:NanotubeDynamics}
	\begin{algorithmic}[1]
		\State Randomly generate a uniform distribution of particle  coordinates $p[n].x$ and $p[n].y$ and angles $p[n].\alpha$, where $n \in [1, 2,\dots, N_p]$.
		\State By default, all particles are marked green.
		\For {$i \leftarrow 1, t_\mathrm{max}/\Delta t$}
		\State Calculate $R_f$
		according to~\eqref{eq:FixingRadiusNanotube}. To avoid dividing by zero
		(at $t \to t_\mathrm{max}$) $R_f \leftarrow 0.5 l$,
		when $R_f < 0.5 l$.
		\For {$n \leftarrow 1, N_p$}
		\State $p[n].moved \leftarrow \texttt{False}$
		\State Calculate the electrostatic forces for the particle $p[n]$ [see formula~\eqref{eq:SumOfForces}].
		\EndFor
		\State $MovementDetected \leftarrow \texttt{True}$
		\State $attempts \leftarrow 0$
		\While {$MovementDetected$}
		\State $MovementDetected \leftarrow \texttt{False}$
		\State $attempts \leftarrow attempts + 1$
		\For {$n \leftarrow 1, N_p$}
		\State Override the color of the particle with regard to its position relative to $R_f$.
		\State Skip the red particles.
		\If {$p[n].moved = \texttt{False}$}
		\If {(the particle is green)}
		\If {$attempts = 1$}
		\State Calculate new coordinates of the particle resulting from the diffusion [see formula~\eqref{eq:DiffusionDisplacementAndRotation}].
		\EndIf
		\State Calculate new coordinates of the particle resulting from the advection [see formulas~\eqref{eq:velocityAnalytical}, \eqref{eq:AdvectionDisplacement}, and \eqref{eq:velocityAnalyticalModified}].
		\State Calculate new coordinates of the particle resulting from the electrostatics and drag forces [see formulas~\eqref{eq:particleVelocityOnFrictionAndElectrostatic} and \eqref{eq:AngularVelocityByViscosityAndElectrostatics}].
		\If {(no collision)}
		\State Move the particle.
		\State $p[n].moved \leftarrow \texttt{True}$
		\State $MovementDetected \leftarrow \texttt{True}$
		\Else
		\State Calculate new coordinates of particle by rotating it against collision point [see formulas~\eqref{eq:RotationAngleByFlowRelativeToEnd}, \eqref{alphaflownotend}, \eqref{eq:AngularVelocityByViscosityAndElectrostaticsAgainstItsEnd}, and \eqref{eq:AngularVelocityByViscosityAndElectrostaticsAgainstArbitraryPoint}].
		\If {(no collision)}
		\State Move the particle.
		\State $p[n].moved \leftarrow \texttt{True}$
		\State $MovementDetected \leftarrow \texttt{True}$
		\EndIf
		\EndIf
		\ElsIf {(the particle is black)}
		\State Calculate new coordinates of the particle resulting from rotation against its intersection point with $R_f$ [see formulas~\eqref{eq:RotationAngleByFlowRelativeToEnd} and \eqref{eq:AngularVelocityByViscosityAndElectrostaticsAgainstArbitraryPoint}].
		\If {(no collision)}
		\State Move the particle.
		\State $p[n].moved \leftarrow \texttt{True}$
		\State $MovementDetected \leftarrow \texttt{True}$
		\EndIf
		\EndIf
		\EndIf
		\EndFor
		\EndWhile
		\State Write the coordinates of the particles and their colors for the current time step to a file.
		\EndFor
	\end{algorithmic}
\end{algorithm}

Due to the impossibility of fulfilling the condition of $\Delta t \to 0$ in the numerical simulations, large angles of rotation of a nanotube can occur due to a strong electrostatic impulse caused by a very close position of the current particle relative to its neighbor. This problem is caused by the discreteness of time. For each small time step, the rotation of the particle must be negligible, as it corresponds to a smooth motion without sudden jumps. This also applies to the flow rotation. For this reason, we check the condition at each time step. If $\alpha > \alpha_\mathrm{max}$, we assign a value $\alpha$ to $\alpha_\mathrm{max}$. For calculations, we chose $\alpha_\mathrm{max}=1^\circ$, having previously compared it to the case when $\alpha_\mathrm{max}=10^\circ$ (no qualitative differences have been noticed). In addition, the numerical evaluation of the rotation angle of the nanotube in the case where the particle is subjected to the maximum torque from the liquid (the nanotube is fixed by its edge at the boundary $R_f$) using formula~\eqref{eq:RotationAngleByFlowRelativeToEnd} has shown that the maximum rotation angle is less than $1^\circ$. Additional discussion of $\alpha_\mathrm{max}$ usage is provided in Supplemental Material~A, Sec. ~3.

Let each nanotube be in one of three states, with a specific color used for each particle depending on its status: movable (green), frontier (black), and motionless (red). At the initial moment, particles are placed randomly in the area to ensure that they do not overlap (without intersections). By default, all nanotubes are colored green, indicating that they are subject to various types of transfer: advection, diffusion, and electrostatic interactions. Those particles that touch the boundary of $R_f$ are colored black. Such particles are subject only to advective and electrostatic forces, since partial contact with the substrate is assumed. Nanotubes beyond the fixing radius ($r > R_f$) are colored red. This color label identifies particles that have established contact with the substrate (deposit).

At each time step, all particles are processed. Each particle undergoes sequentially the transfer mechanisms allowed according to its status. If a collision occurs during particle motion (i.e., if it intersects with at least one of its neighboring particles), the displacement will be rejected in the current time step. The algorithm assumes that particles which be rejected are recorded in order to make further displacement attempts. These attempts do not apply to diffusion, which corresponds to the Monte Carlo method (a single attempt of random displacement/rotation). Such attempts are repeated until one of two events occurs: 1) all particles are displaced, or 2) no particle is displaced during a single pass through the recorded particles.

We consider only pairwise interactions between the target particle and its neighbors within a cutoff radius $R_\mathrm{cut} = 3l$. In reality, such interaction is long-range. But at a distance of $3l$, the interaction force between two point charges is almost an order of magnitude less than at a distance of $l$ (the effect of $R_\mathrm{cut}$ on the simulation results is discussed in Supplemental Material~A, Sec. ~3).

\section{Results and discussion}
Here, to evaluate the ordering of structures, we use the nematic order parameter $S = \left\langle \cos(2\Theta) \right\rangle$, where $\Theta$ is the angle between the nanotube and the nematic direction~\cite{Sanchez2015,Wang2022}. In our case (circular domain), it is the angle between the nanotube and the perpendicular to the radius drawn from the center of the region to the center of the particle~\cite{Sanchez2015} (for details, see the link \href{https://isanm.space}{https://isanm.space}). A value of $S=1$ indicates that the nanotubes are aligned parallel to the external circular boundary. At $S=-1$, particles are aligned perpendicular to this boundary. The value of $S=0$ indicates a chaotic structure. Each computational experiment has been repeated 10 times and then the results have been averaged. Furthermore, the error has been assessed using the standard error of the mean. By default, all three factors (advection, diffusion, and electrostatics) are accounted for in the calculations, unless otherwise stated.

\begin{figure}
 	\begin{minipage}{0.5\linewidth}
 		\center{\includegraphics[width=0.99\linewidth]{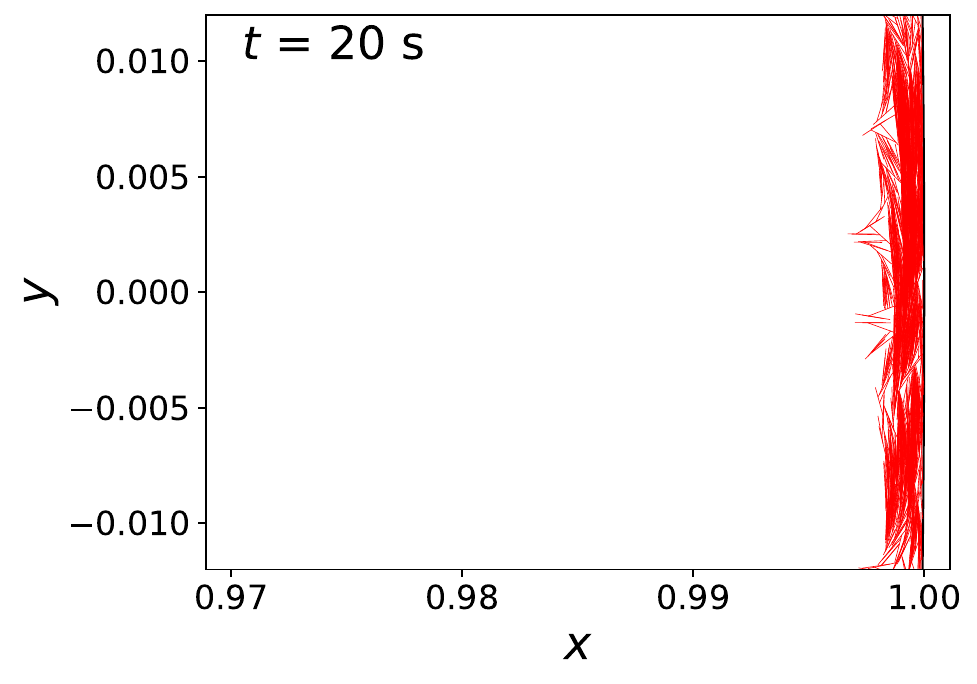} \\ (a)}\\
 		\center{\includegraphics[width=0.99\linewidth]{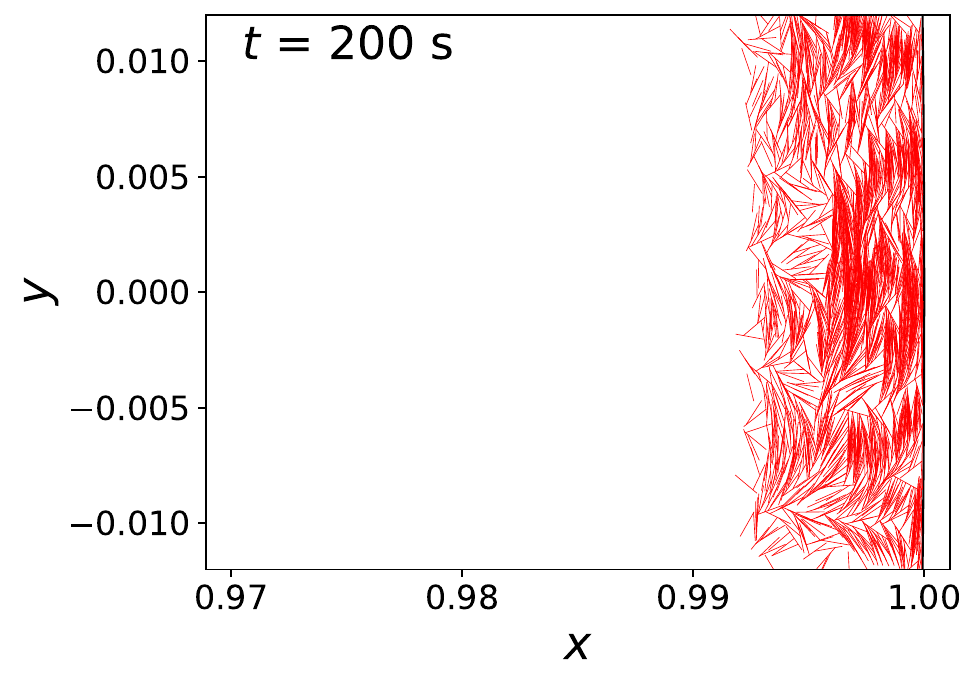} \\ (b)}\\
 		\center{\includegraphics[width=0.99\linewidth]{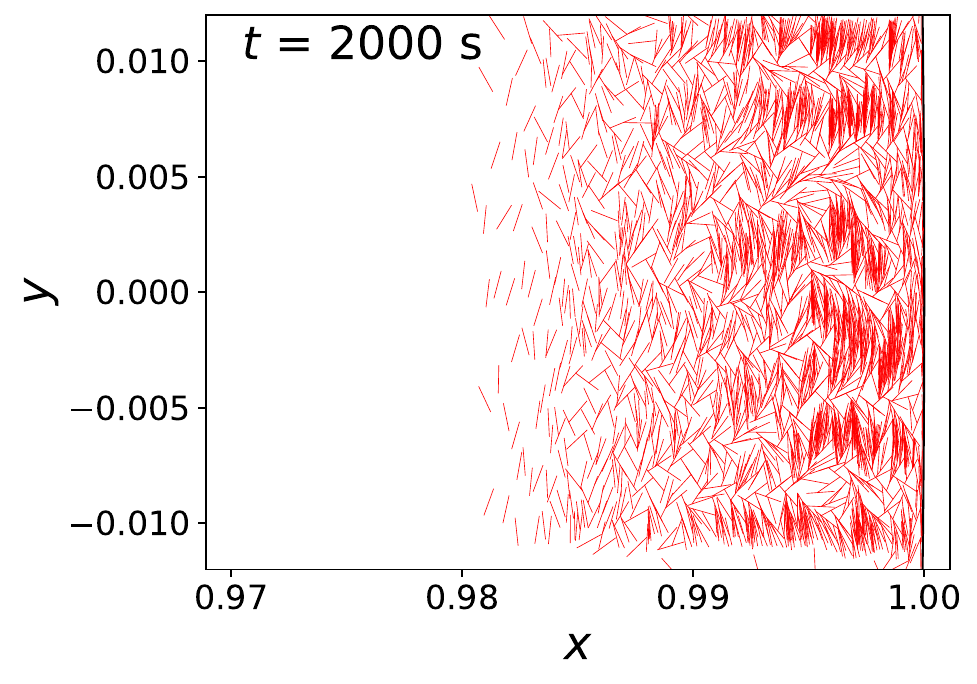} \\ (c)}	
 	\end{minipage}
 	\hfill
 	\begin{minipage}{0.49\linewidth}
 		\center{\includegraphics[width=0.99\linewidth]{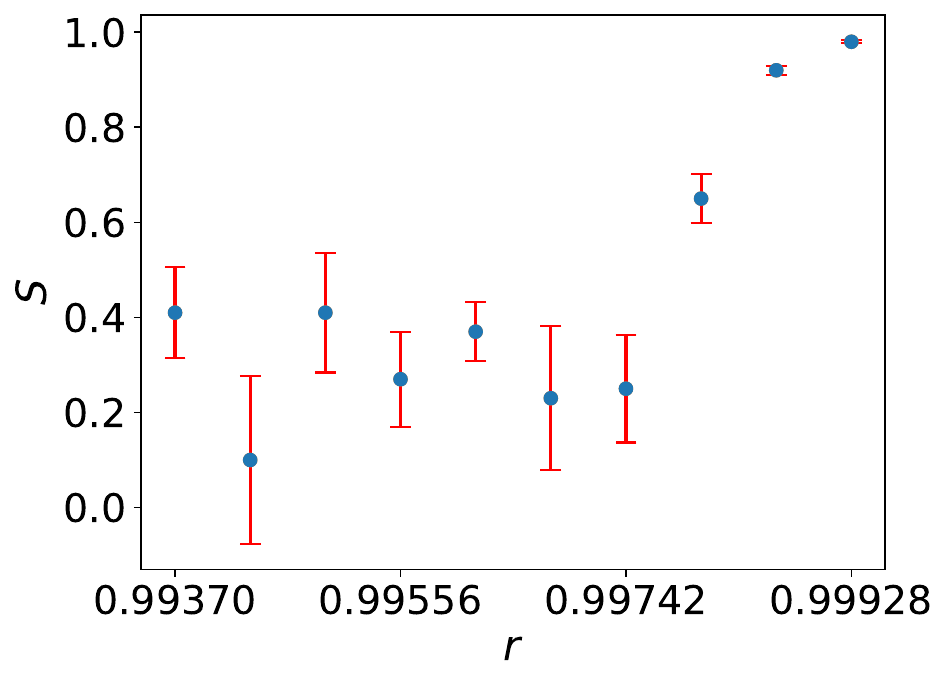} \\ (d)}\\
 		\center{\includegraphics[width=0.99\linewidth]{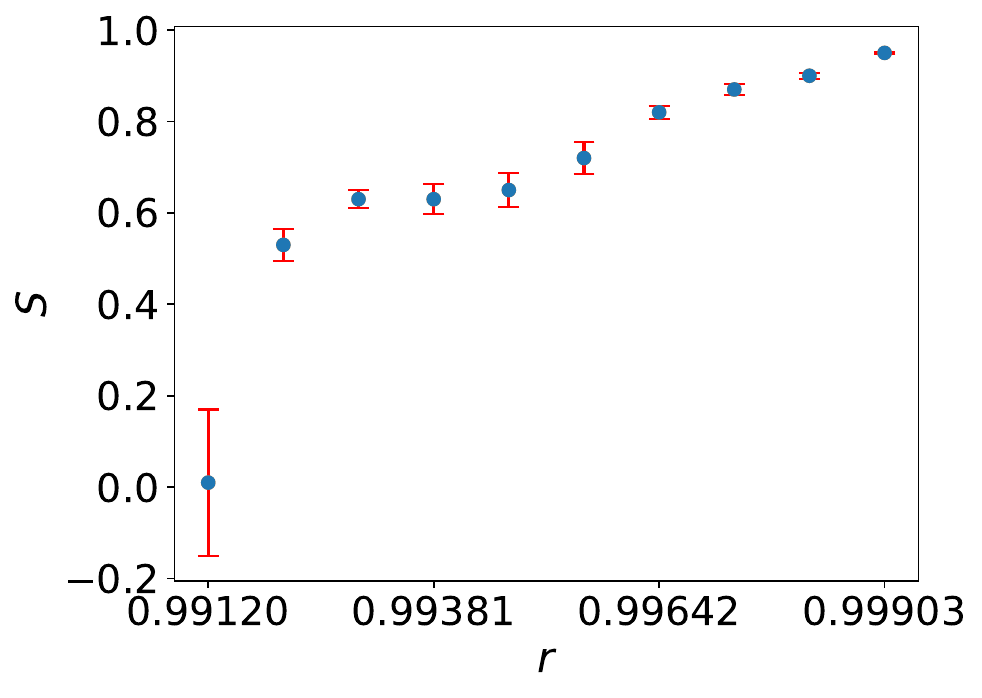} \\ (e)}\\
 		\center{\includegraphics[width=0.99\linewidth]{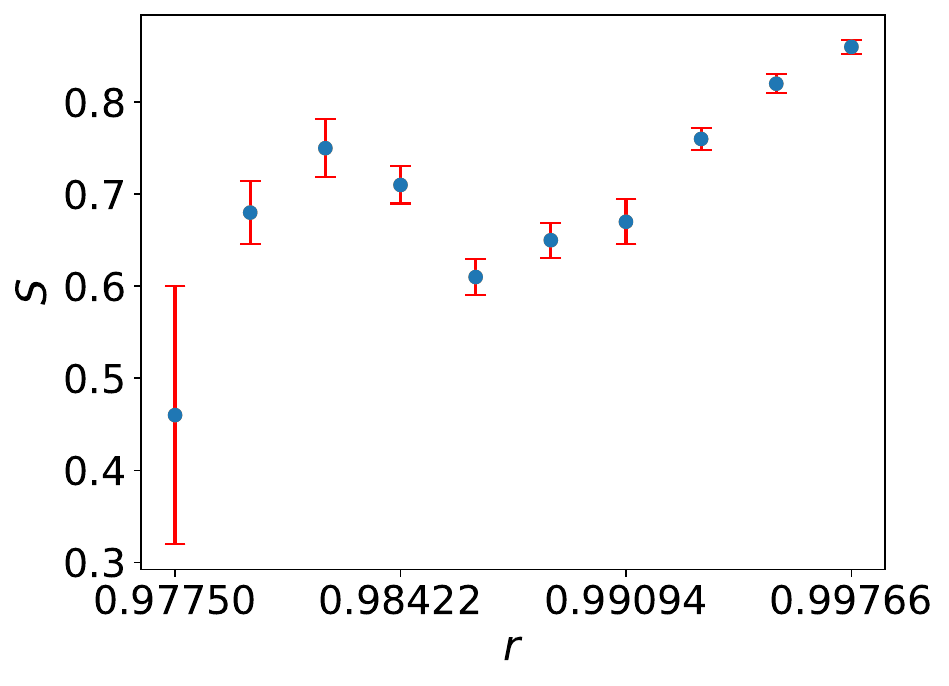} \\ (f)}
 	\end{minipage}
 	\caption{Calculation results taking into account advection, diffusion, and electrostatics for $N_p = 3000$ ($kq^2 = 10^{-26}$ m$^2$N): (a)--(c) deposit structure and (d)--(f) order parameter $S$ at $t_\mathrm{max}=$ 20 s (a, d), 200 s (b, e), and 2000 s (c, f).}
	 \label{fig:AdvDifEl_26_20_200_2000s}
 \end{figure}

According to formula~\eqref{eq:velocityAnalytical}, $v_r \sim
R/t_\mathrm{max}$, meaning that the smaller the droplet evaporation time, the greater the capillary flow velocity, also known as compensatory flow or evaporation-induced flow. A high flow velocity results in a denser packing of nanotubes, as particles near the periphery are pressed to each other more strongly by the flow compared to a slow flow (Fig.~\ref{fig:AdvDifEl_26_20_200_2000s}). Electrostatic repulsion of nanotubes counteracts this compression more noticeably in the case of a slow flow. Increasing the parameter $t_\mathrm{max}$ allows one to obtain a wider deposit layer near the periphery. When the analysis is extended from a specified sector to the entire domain, the deposition pattern resolves into a ring-shaped one. Thus, it becomes possible to control the thickness of the nanotube ring.  With a relatively strong particle charge ($kq^2 = 10^{-26}$ m$^2$N) and the time of $t_\mathrm{max}=$ 200~s, we observe a transition from a chaotic region to an ordered one (moving from the center to the periphery), similar to the experiment (see Fig. 3 in Ref.~\cite{Zhao2015}). In the study~\cite{Zhao2015}, the nanotubes had a negative charge, but this sign is omitted here as it does not affect the result.

\begin{figure}
	\begin{minipage}{0.5\linewidth}
		\center{\includegraphics[width=0.99\linewidth]{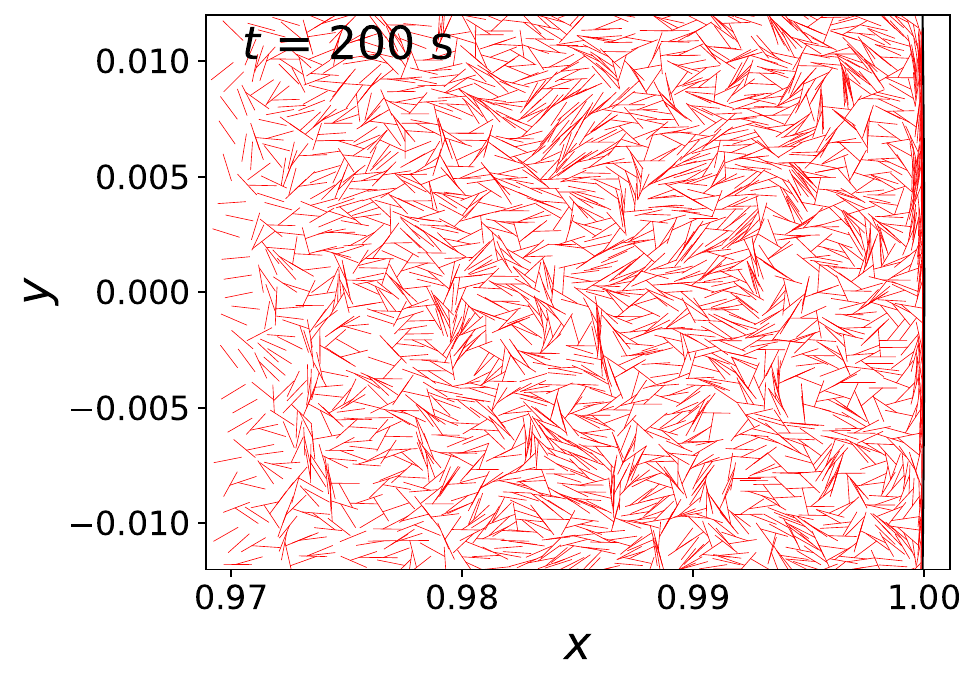} \\ (a)}\\
		\center{\includegraphics[width=0.99\linewidth]{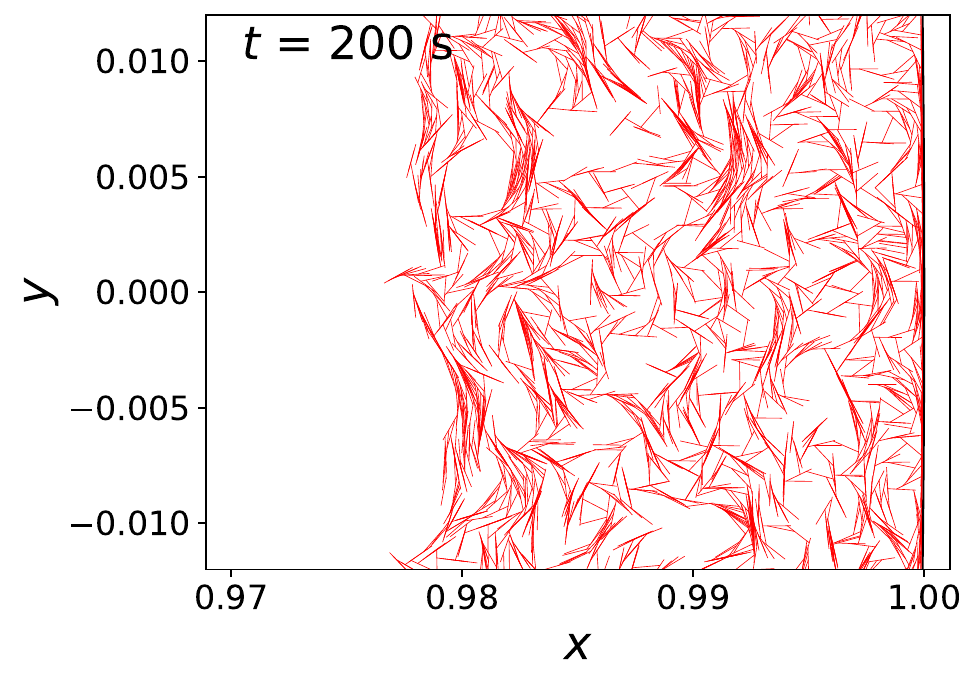} \\ (b)}	
	\end{minipage}
	\begin{minipage}{0.49\linewidth}
		\center{\includegraphics[width=0.99\linewidth]{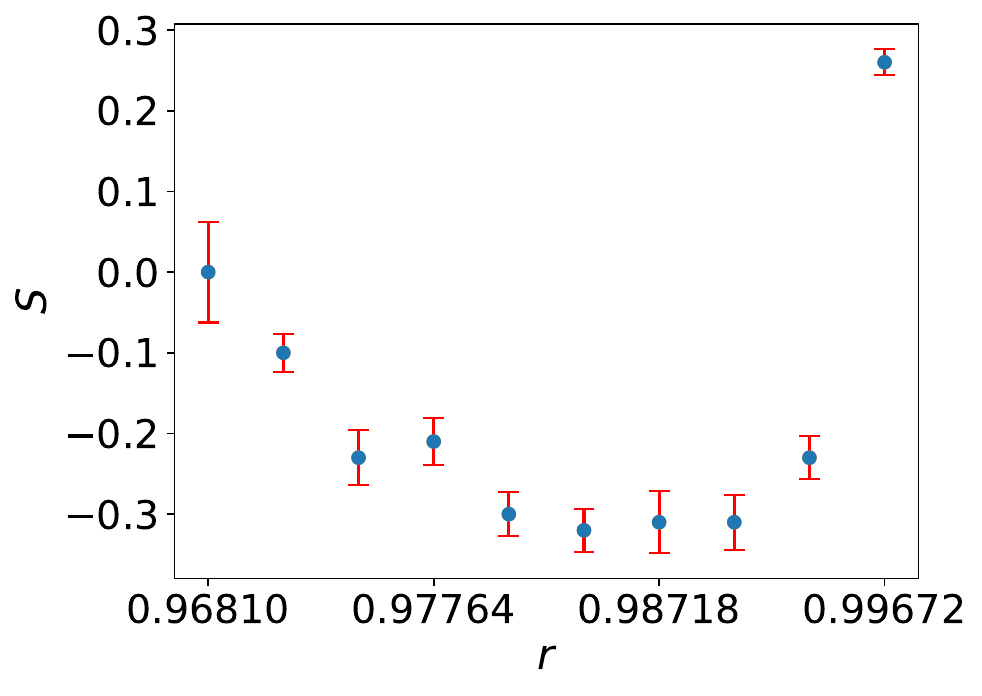} \\ (c)}\\
		\center{\includegraphics[width=0.99\linewidth]{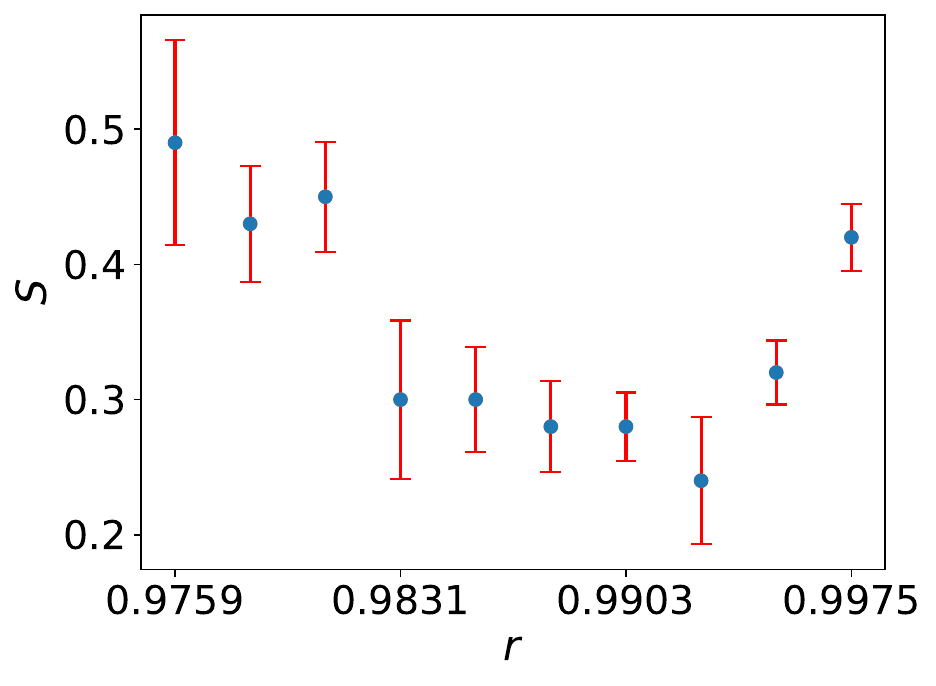} \\ (d)}
	\end{minipage}
	\caption{Calculation results for $N_p = 3000$ and $t_\mathrm{max}=$ 200~s considering only advection and electrostatics: (a),(b) deposit structure and (c),(d) order parameter $S$ for $kq^2 = 10^{-26}$ m$^2$N (a),(c) and $kq^2 = 10^{-28}$ m$^2$N (b),(d).}
	\label{fig:AdvEl_200s}
\end{figure}

Without taking into account diffusion, the model predicts the chaotic structure (Fig.~\ref{fig:AdvEl_200s}). There is no transition to order in this case. This indicates that diffusion contributes by allowing the nanotubes to arrange more densely. With a relatively low charge ($kq^2 = 10^{-28}$ m$^2$N) many local voids are observed in the structure [Fig.~\ref{fig:AdvEl_200s}(b)]. The shape of some of them is close to that of a circle, with a diameter several times the length of the nanotube. Such voids were not observed in the experiment~\cite{Zhao2015}. The strong electrostatic interaction induces prealignment of nanotubes prior to deposition. As a result, such noticeable voids are not formed [Fig.~\ref{fig:AdvEl_200s}(a)].

\begin{figure}
	\begin{minipage}{0.5\linewidth}
		\center{\includegraphics[width=0.99\linewidth]{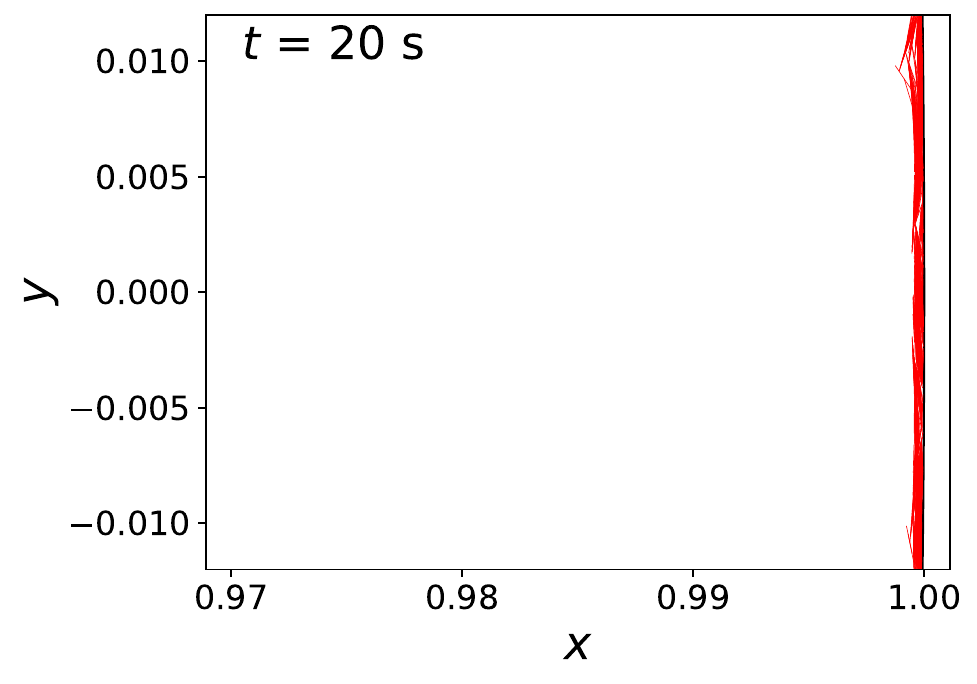} \\ (a)}\\
		\center{\includegraphics[width=0.99\linewidth]{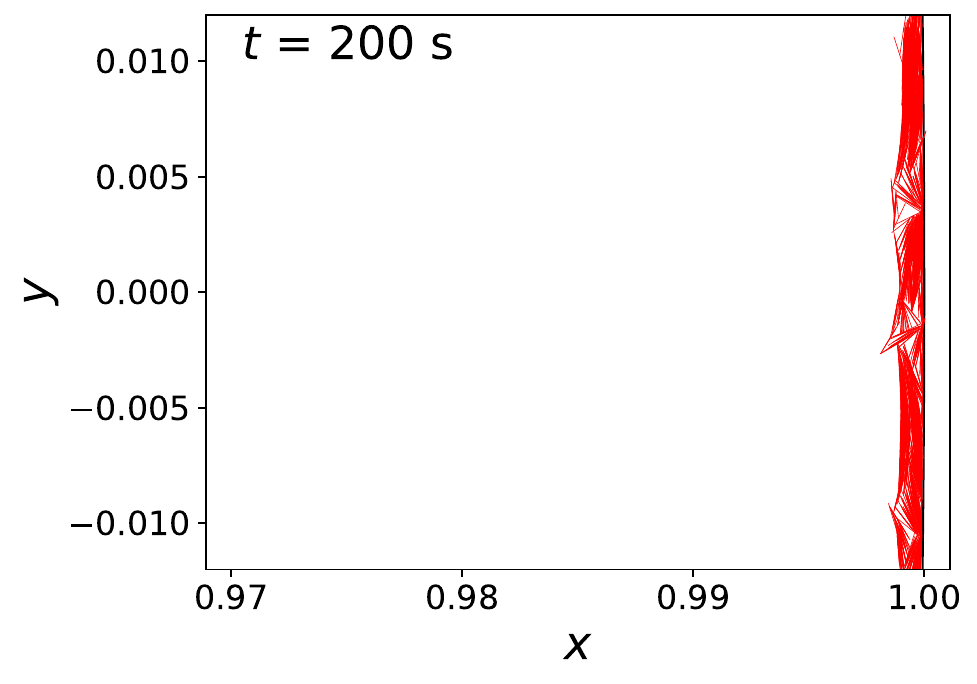} \\ (b)}\\
		\center{\includegraphics[width=0.99\linewidth]{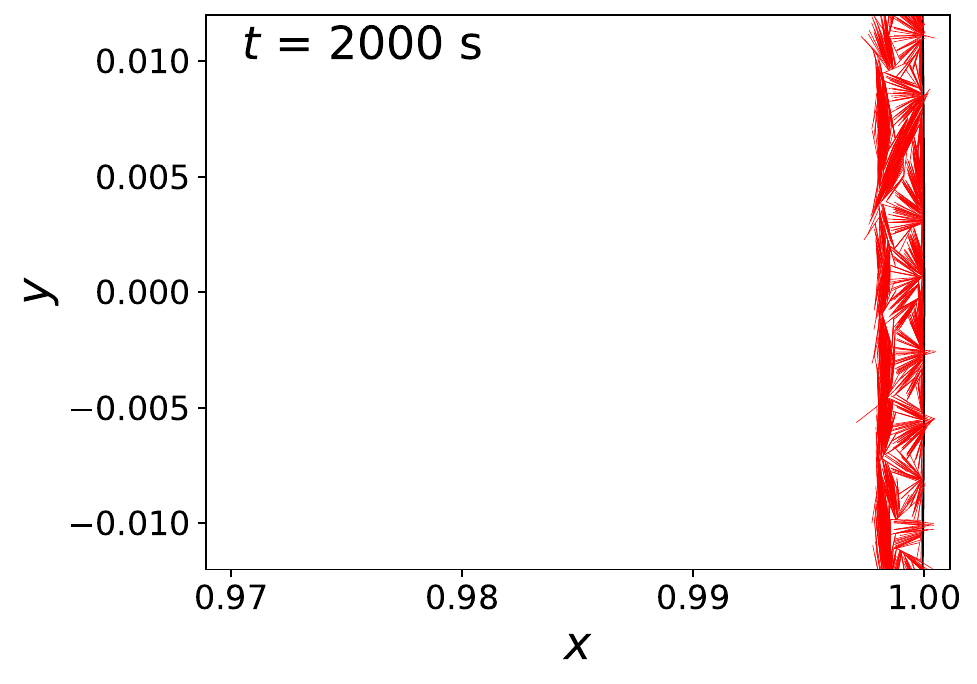} \\ (c)}	
	\end{minipage}
	\begin{minipage}{0.49\linewidth}
		\center{\includegraphics[width=0.99\linewidth]{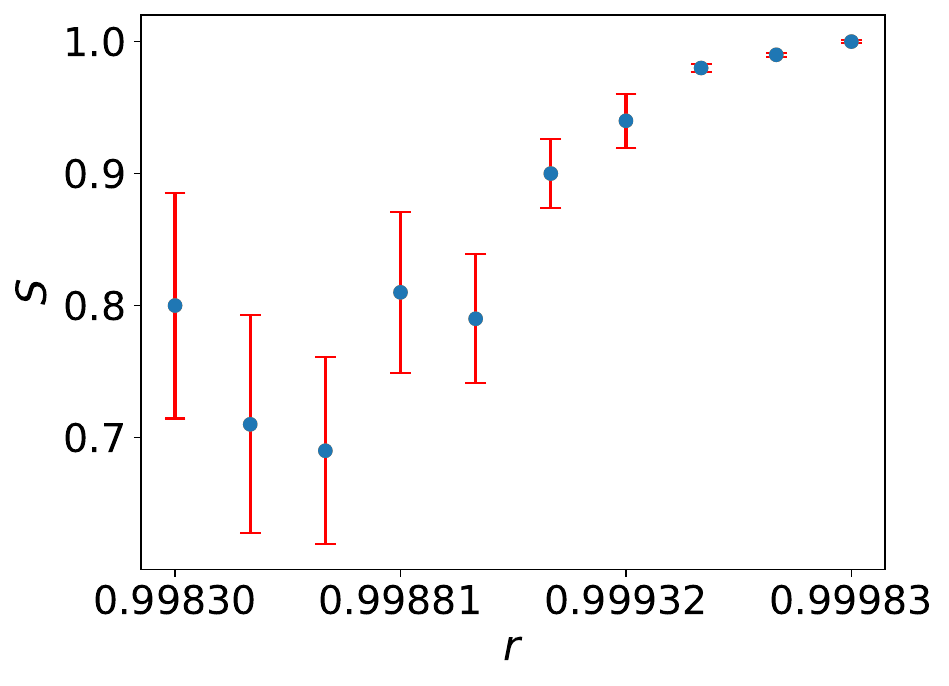} \\ (d)}\\
		\center{\includegraphics[width=0.99\linewidth]{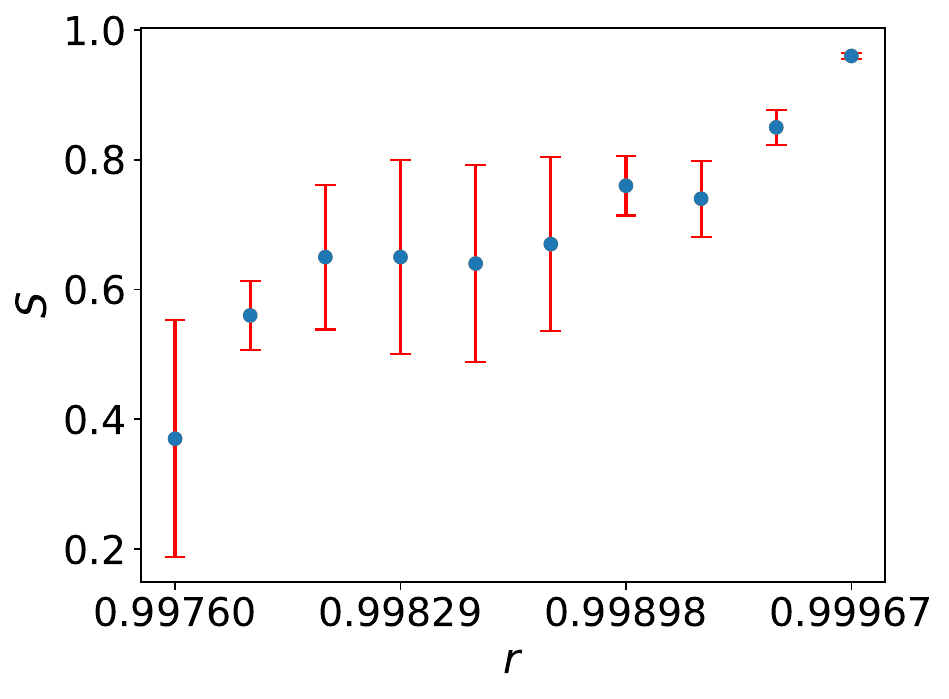} \\ (e)}\\
		\center{\includegraphics[width=0.99\linewidth]{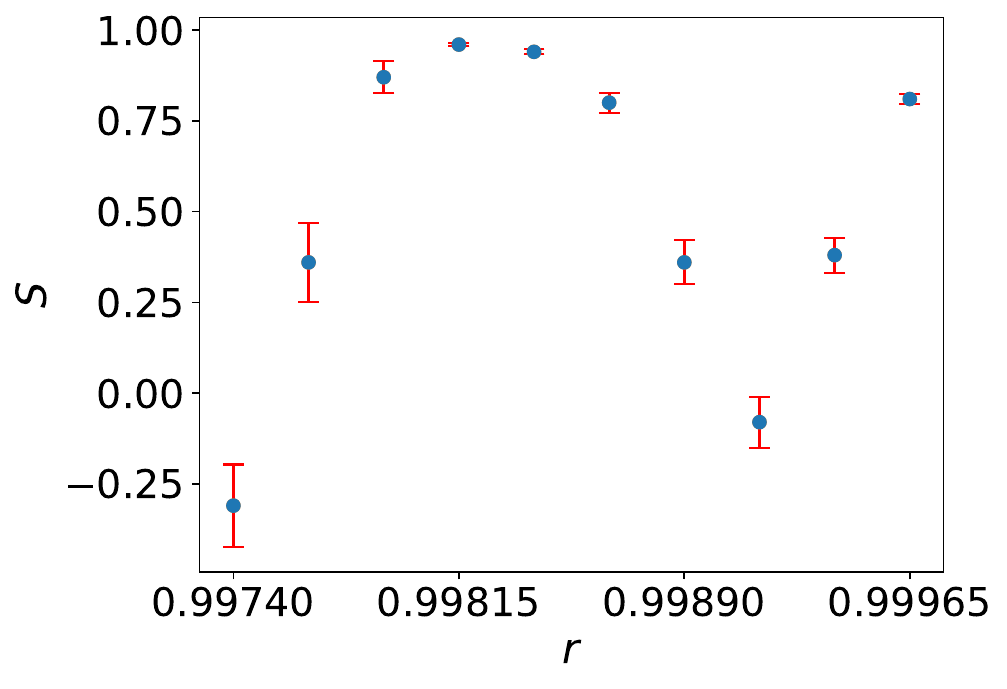} \\ (f)}
	\end{minipage}
	\caption{The calculation results for $N_p = 3000$ taking into account advection, diffusion, and electrostatics ($kq^2 = 10^{-28}$ m$^2$N): (a)--(c) deposit structure and (d)--(f) order parameter $S$ for $t_\mathrm{max}=$ 20 s (a),(d), 200 s (b),(e) and 2000 s (c),(f).}
	\label{fig:AdvDifEl_28_20_200_2000s}
\end{figure}

With a small charge ($kq^2 = 10^{-28}$ m$^2$N), the structure turns out to be more dense (Fig.~\ref{fig:AdvDifEl_28_20_200_2000s}), than with a large one ($kq^2 = 10^{-26}$ m$^2$N), see Fig.~\ref{fig:AdvDifEl_26_20_200_2000s}. It means that the morphology of the deposit is also determined by the confrontation between advection and electrostatics, as it has been already noted above in the text. With decreasing $t_\mathrm{max}$ or $kq^2$, advection begins to dominate electrostatics. In the case of an increase in the values of these parameters, the influence of electrostatics becomes more noticeable. As a result, advection fades into the background. For this reason, the width of the nanotube layer becomes thinner (Fig.~\ref{fig:AdvDifEl_28_20_200_2000s}).
Note that nonlinear behavior of $S$ is observed at $t_\mathrm{max}=$ 2000 s [Figs.~\ref{fig:AdvDifEl_28_20_200_2000s}(c) and \ref{fig:AdvDifEl_28_20_200_2000s}(f)]. Such layered morphology qualitatively resembles the results of other experimental work (see Fig.~6~\cite{Denneulin2011}).

\begin{figure}
	\center{\includegraphics[width=0.8\linewidth]{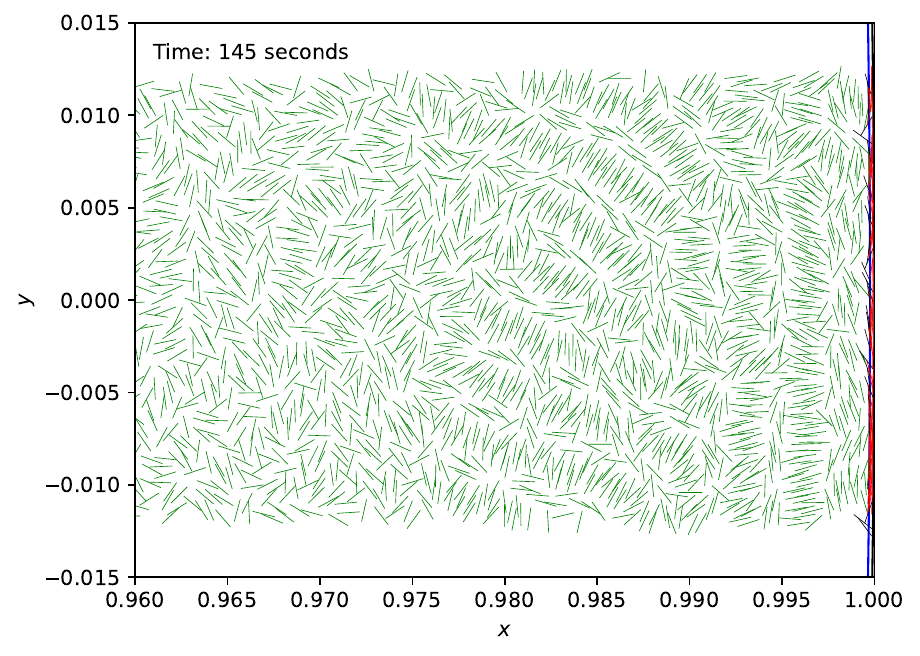}(a)}\\	
	\center{\includegraphics[width=0.8\linewidth]{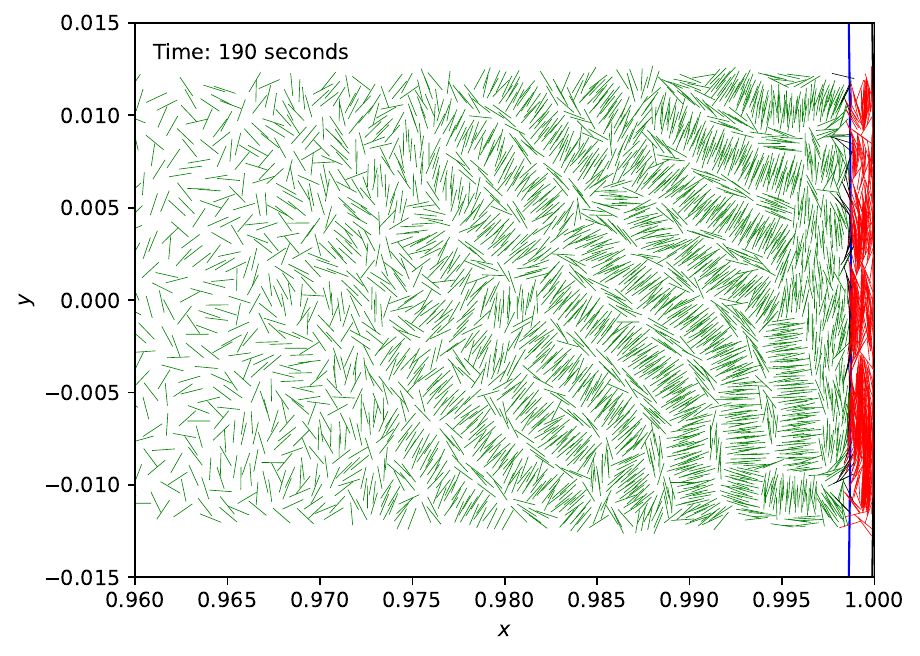}(b)}	
	\caption{Formation of stack structures at different points of evaporation time: (a) $t=$ 145~s and (b) $t=$ 190~s (calculation parameters: $N_p=4000$, $kq^2=10^{-26}$ m$^2$N, $t_\mathrm{max}=200$ s).}
	\label{fig:stacks}
\end{figure}

When considering diffusion, with increasing concentration and charge, the formation of local ordered stack structures (Fig.~\ref{fig:stacks}) becomes more pronounced, gradually arising, moving and disappearing (see Supplemental Material~B). Similar structures were previously observed in a lattice model describing self-organization in a 2D system of rods with two mutually perpendicular orientations using random walk~\cite{Ulyanov2018}. The fixing radius is indicated by a blue line in Fig.~\ref{fig:stacks}.

\begin{figure}
	\includegraphics[width=0.85\linewidth]{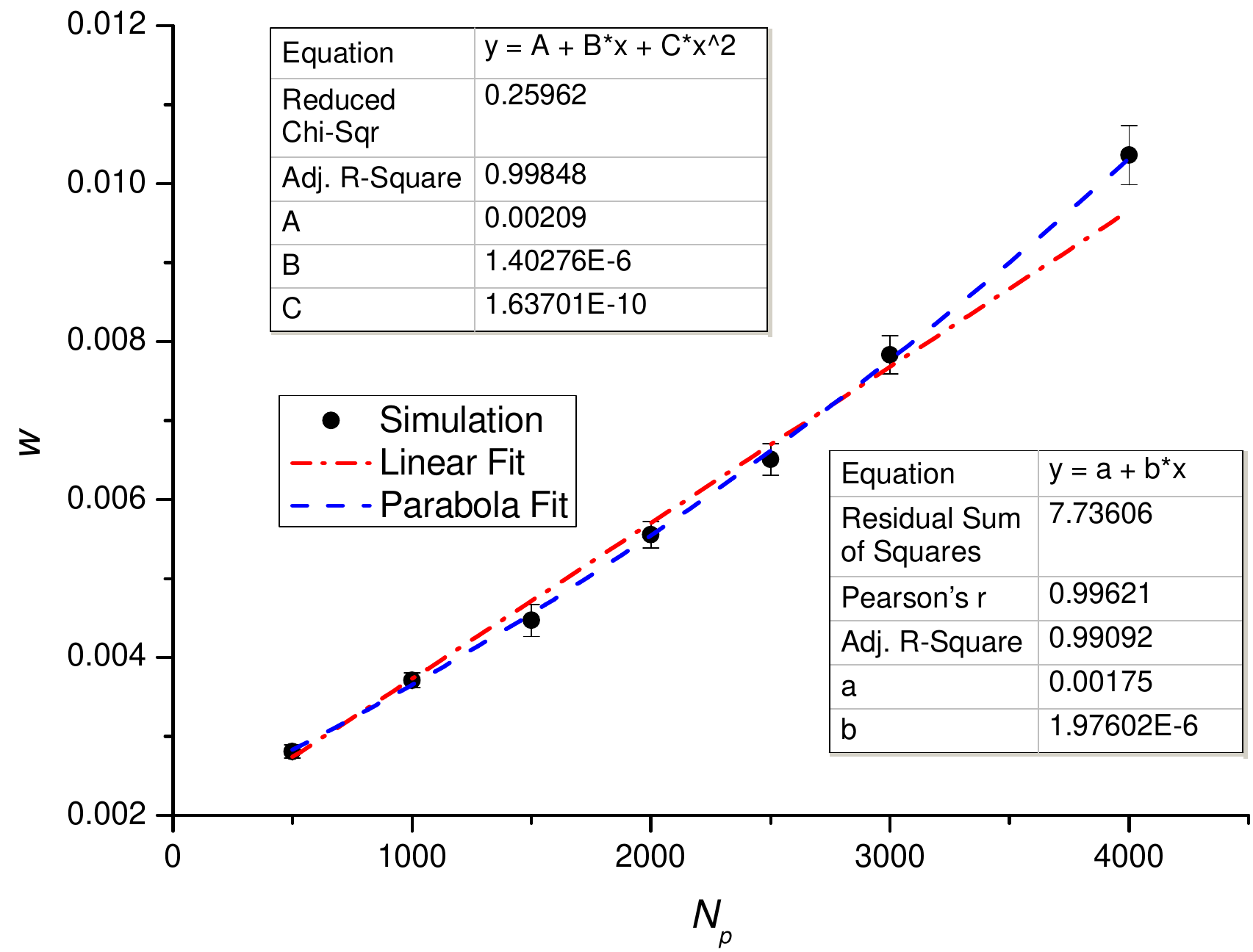}
	\caption{\label{fig:depositWidthVsParticleNumber}Dependence of the deposit width on the number of particles.}
\end{figure}

There are no qualitative differences in the order parameter, $S$, observed for different values $N_p$ (see Supplemental Material~A, Sec. 2). In the experiment~\cite{Zhao2015}, it was shown that, at a low concentration of the solution, alignment of the nanotubes was not observed. The proposed 2D model does not allow this observation to be reproduced. To do this, it is necessary to further develop a 3D model with a multilayer deposit. The 2D model presented here predicts that the width of the deposit, $w$, grows with increasing initial concentration of the solution. There is a parabolic dependence of $w$ on $N_p$ (Fig.~\ref{fig:depositWidthVsParticleNumber}).
It should be noted that, in the results presented here, $x$, $y$, and $w$ are scaled by $R$.

\section{Conclusion}
The results of the numerical simulations indicate that the combined effect of advection, diffusion, and electrostatics, under certain values of key system parameters ($N_p$, $t_\mathrm{max}$, and $k q^2$) leads to a transition from order to disorder in the morphology of the nanotube ring deposit, associated with the coffee-ring effect, which was previously observed in the experiment~\cite{Zhao2015}. The concentration of particles affects the average distance between nanotubes in a solution. The strength of the electrostatic interaction between the nanotubes depends on the distance between them and the magnitude of their charge. This interaction can induce mutual nanotube alignment during the initial stages of the process. The evaporation time affects the velocity of capillary flow, which transfers particles and rotates them when they reach the periphery of a droplet, where the thickness of the liquid layer corresponds to the diameter of nanotubes. This boundary is known as the fixing radius, $R_f$~\cite{Kolegov2019}. Each nanotube that crosses this boundary is deposited on the substrate (i.e., it comes into contact with the substrate and sticks to it due to adhesion). In the model, the boundary $R_f$  moves as the liquid evaporates until it collapses at the central point of the region. In reality, the situation may be more complicated due to the appearance of many local ruptures (dry spots) in an ultrathin film. However, this does not affect the results of the numerical simulations presented here, as such local ruptures, if they occur, appear only after the ring-shaped deposit has been formed. For further research, a modified method with restricted evaporation is also interesting, when a droplet is located between two substrates~\cite{Jiang2022}. This experiment showed the possibility of obtaining aligned nanotube precipitation with an area of several square centimeters. In addition, it is necessary to develop 3D models for a more accurate description of these processes. In certain cases, additional effects must be taken into account, such as sedimentation, contact line motion, and hydrodynamic and capillary interactions between particles~---~including those in binary mixtures with varying shapes (e.g. spheres and ellipsoids~\cite{Yunker2011}) or sizes.

\begin{acknowledgments}
This work is supported by Grant No. 22-79-10216 from the Russian Science Foundation (\href{https://rscf.ru/en/project/22-79-10216/}{https://rscf.ru/en/project/22-79-10216/}).
\end{acknowledgments}

\section*{Data Availability Statement}

The data that support the findings of this study are
available from the corresponding author upon reasonable
request.

\appendix

\section{Rotation of the nanotube by the fluid flow}\label{appendix:FlowRotation}

Consider the rotation relative to the center of the nanotube mass at point $C$.
The moment of the force exerted on a particle by a liquid relative to a given point is equal to
\begin{equation}
\vec{M} = \int_{A_{1}A_{2}} \left[\overrightarrow{CA}, \, dm \left.\frac{d\vec{v}}{dt}\right|_{A}\right].
\label{eqAone}
\end{equation}
 Here, $A_{1}$ and $A_{2}$ are the nanotube edges,  $A$ is some point of the nanotube, $dm$ is the mass of an elementary piece of the nanotube, and $dl$ is the length of this piece ($dm = dl\,m/l$).

Taking into account formula~\eqref{eq:velocityAnalytical} for the flow velocity averaged over the height of the droplet, it is possible to determine the average acceleration of the liquid in projections onto the coordinate axes:
\begin{equation}\label{eqaccelerationx}
\frac{dv_{x}}{dt} =\frac{R^{2}x}{4 \left(x^{2} + y^{2}\right) \left(t_\mathrm{max} - t   \right)^{2}} \left(\frac{1}{\sqrt{1 - \frac{x^{2} + y^{2}}{R^{2}}}} - 1 + \frac{x^{2} + y^{2}}{R^{2}}    \right),
\end{equation}
\begin{equation}\label{eqaccelerationy}
\frac{dv_{y}}{dt} =\frac{R^{2}y}{4 \left(x^{2} + y^{2}\right) \left(t_\mathrm{max} - t   \right)^{2}} \left(\frac{1}{\sqrt{1 - \frac{x^{2} + y^{2}}{R^{2}}}} - 1 + \frac{x^{2} + y^{2}}{R^{2}}   \right).
\end{equation}

Expanding the cross product of the vectors in (\ref{eqAone}) and substituting (\ref{eqaccelerationx}) and (\ref{eqaccelerationy}), we obtain

\begin{multline}
\label{eqAtwo}
\vec{M} = \int_{A_{1}A_{2}} \vec{n}_{z} \frac{m}{l}  \left(\left(x_{A} - x_{C}  \right)\left.\frac{d v_{y}}{dt}\right|_{A} -  \left(y_{A} - y_{C}  \right)\left.\frac{d v_{x}}{dt}\right|_{A}  \right) dl =\\
= \vec{n}_{z} \frac{\frac{m}{l} R^{2}}{4 \left(t_\mathrm{max} - t   \right)^{2}} \int_{A_{1}A_{2}} \frac{ y_{A} \left(x_{A} - x_{C}  \right)  - x_{A} \left(y_{A} - y_{C}  \right) }{ x_{A}^{2} + y_{A}^{2}}\times\\ \times\left(\frac{1}{\sqrt{1 - \frac{x_{A}^{2} + y_{A}^{2}}{R^{2}}}} - 1 + \frac{x_{A}^{2} + y_{A}^{2}}{R^{2}} \right) dl.
\end{multline}

The position of the particle is determined by the coordinates of its center of mass $(x_{C},\,y_{C})$ and the angle $\alpha$ between the nanotube and the $OX$ axis. Let us make the following substitutions:
$y_{A} = \tan (\alpha)\times x_{A} + b, \quad dl = d x_A / \cos(\alpha), \quad \text{where}\quad b = y_{C} - \tan (\alpha) \times x_{C}$, and compute the integral,
\begin{multline}
\label{eqAthree}
\vec{M} = \frac{\vec{n}_{z} R^{2} m/ l}{4 \left(t_\mathrm{max} - t   \right)^{2}}
\int_{x_{A_{1}}}^{x_{A_{2}}} \frac{b(x_A-x_C)}{\left(\frac{x_A}{\cos(\alpha)}+ b \sin(\alpha)\right)^{2} + b^2 \cos^2 (\alpha)} \times \\ \times \left(\frac{1}{\sqrt{1-\frac{\left(x_A/ \cos(\alpha) + b\sin(\alpha)\right)^2 + b^2 \cos^2(\alpha)}{R^2}}} - 1\right. +\\
+\left. \frac{\left( \frac{x_{A}}{\cos (\alpha)}+ b \sin (\alpha)\right)^{2} + b^{2} \cos ^{2} (\alpha)}{R^{2}} \right) \frac{d x_A}{\cos(\alpha)} = \frac{\vec{n}_{z} R^2 b\, m/l}{4 \left(t_\mathrm{max} - t   \right)^{2}} \times \\ \times
\left[\cos (\alpha)\left( f_1(g_{A1}) - g_{A1}+\ln(g_{A1}) - f_1(g_{A2}) + g_{A2}-\ln (g_{A2}) \right)\right. -\\
-  \frac{b \sin(\alpha) \cos (\alpha) +  x_{C}}{R \sqrt{W}}
  \left\{\arctan \left(\frac{0.5(W g_{A2}-2W+g_{A2}) }{\sqrt{W\left(1-g_{A2}  \right)\left(g_{A2}-W  \right)}}    \right) \right.+ \\
    + f_2 (g_{A2})  - 2 \arctan \left(\frac{\sqrt{g_{A2}-W}}{\sqrt{W}}  \right)
  -  \\ - \arctan \left(\frac{W g_{A1}-2W+g_{A1} }{2 \sqrt{W\left(1-g_{A1}  \right)\left(g_{A1}-W  \right)}}    \right) - \\
   - \left.\left. f_2 (g_{A1})  + 2 \arctan \left(\frac{\sqrt{g_{A1}-W}}{\sqrt{W}}  \right)     \right\}      \right],
\end{multline}

where $$g_{A1} = \frac{\left(x_{A1}/\cos(\alpha)+ b \sin(\alpha)\right)^{2} + b^2 \cos^2 (\alpha)}{R^2},$$
 $$g_{A2} = \frac{\left(x_{A2}/\cos(\alpha) + b\sin(\alpha)\right)^2 + b^2 \cos^2 (\alpha)}{R^2},$$
$$W = \left( \frac{b \cos(\alpha)}{R}\right)^2,$$
$$f_1(\xi) = 2\arctanh \sqrt{1-\xi},$$
$$f_2(\xi) = 2 \sqrt{W} \sqrt{\xi-W},$$
$\xi$ is $g_{A1}$ or $g_{A2}$.

The torque can also be approximated by using~\eqref{eqAone}:
\begin{multline}
\label{eqAfour}
\vec{M} =  \int_{A_1 C} \left[\overrightarrow{CA}, \, dm \left.\frac{d\vec{v}}{dt}\right|_{A}\right] +  \int_{C A_2} \left[\overrightarrow{CA}, \, dm \left.\frac{d\vec{v}}{dt}\right|_{A}\right]
\approx \\ \approx \left[\overrightarrow{CS_{1}}, \,  \left.\frac{d\vec{v}}{dt}\right|_{S_{1}}\right] \frac{m}{2} + \left[\overrightarrow{CS_{2}}, \, \left.\frac{d\vec{v}}{dt}\right|_{S_{2}}\right]  \frac{m}{2}
= \\ = \vec{n}_{z} \frac{m R^2}{8 \left(t_\mathrm{max} - t   \right)^2}
\left( f_3(x_{S1},y_{S1})  +  f_3(x_{S2},y_{S2})    \right),
\end{multline}
where $S_1$ and $S_2$ are some points on the nanotube located on opposite sides of its center,
\begin{multline*}
\hfill f_3(x,y) = \hfill \\ 
= \frac{y\left(x - x_C\right) - x\left(y - y_C\right)}{x^2 + y^2} 
\left(\frac{1}{\sqrt{1 - \frac{x^2 + y^2}{R^2}}} - 1 + \frac{x^2 + y^2}{R^2}\right).
\end{multline*}

\begin{figure}
\centering
\includegraphics[width=.5\columnwidth]{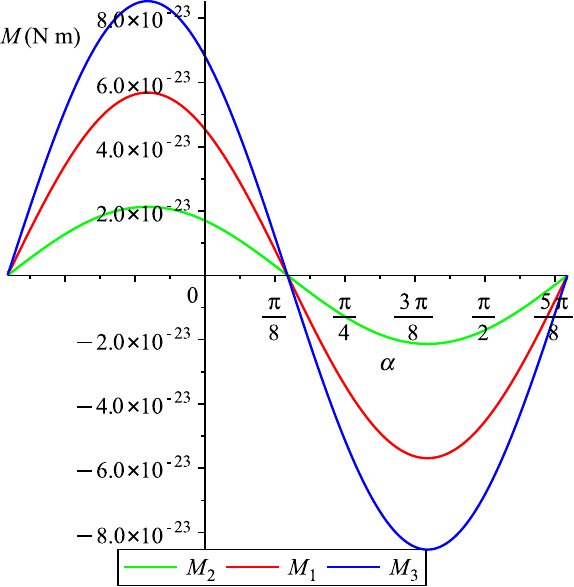}
\caption{Dependence of the torque acting on a particle from the liquid, relative to the center of the nanotube, on the angle $\alpha$ ($x_C=10^{-3}$~m, $y_C=0.5 \times 10^{-3}$~m).}
\label{figAone}
\end{figure}

\begin{figure}
\centering
\includegraphics[width=.5\columnwidth]{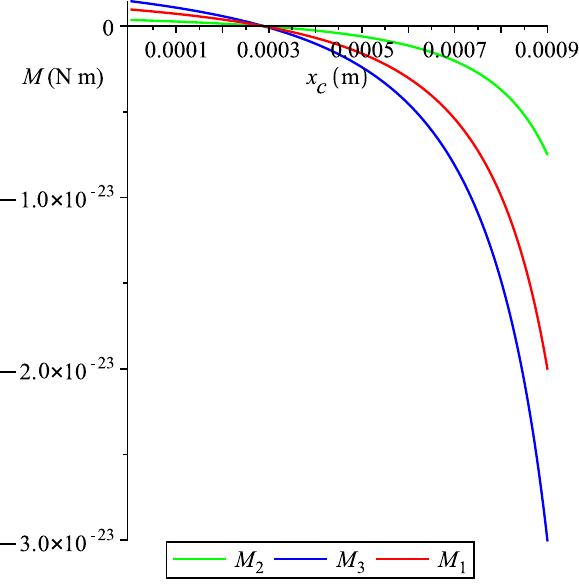}
\caption{Dependence of the torque acting on a particle from the liquid, relative to the center of the nanotube, on $x_{C}$ coordinate of the nanotube ($\alpha = \pi/3$,  $y_C=0.5 \times 10^{-3}$~m).}
\label{figAtwo}
\end{figure}

A comparative analysis of the formulas for the torque acting on a particle from the liquid relative to the center of a nanotube is presented in Figs. \ref{figAone} and \ref{figAtwo} for the time moment $t=100$~s. Note that $t_\mathrm{max}= 200$~s, $M_{1}$ is calculated by the formula~\eqref{eqAthree}, $M_{2}, \, M_{3}$ are calculated by the formula~\eqref{eqAfour}. For points $S_{1}$ and  $S_{2}$ in $M_{2}$,  the midpoints of the nanotube halves (points $E$ and $G$) have been used; in $M_{3}$, two points at the ends of the nanotube ($A_{1}$ and $A_{2}$) have been used.
The torque calculated using the approximate formula~\eqref{eqAfour} exhibits fairly good accuracy and can be employed for calculations in place of the exact formula~\eqref{eqAthree}, which slows down the computation process.

\begin{figure}
\centering
\includegraphics[width=.5\columnwidth]{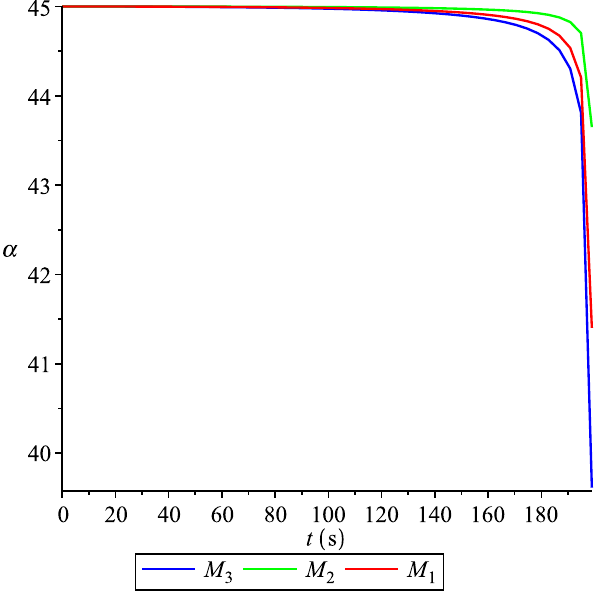}
\caption{Dependence of the angle $\alpha$ (in degrees) on time. Initial conditions: $\left. x_C\right|_{t=0} = 10^{-4}$~m, $\left. y_C\right|_{t=0} = 0$, $\left. \alpha \right|_{t=0} = 45^\circ$, the initial angular velocity is zero (the parameter $t_\mathrm{max}= 200$~s).}
\label{figAthree}
\end{figure}

The rotation angle of the nanotube during particle motion in liquid, while the fixing radius has not been reached, can be determined by solving the system of differential equations:
\begin{equation}
\label{eqAfive}
\frac{1}{12} m l^{2} \frac{d^{2} \vec{\alpha}}{d t^{2}} = \vec{M}, \qquad \frac{d x_{C}}{d t} = v_{x}, \qquad  \frac{d y_{C}}{d t} = v_{y}.
\end{equation}
Figure ~\ref{figAthree} shows the result of a numerical calculation of the change in angle $\alpha$ (in degrees) over time. The calculation was carried out until the particle reached the fixing radius. The calculation results for different initial positions of the particle show that the rotation of the nanotube begins near the edge of the droplet, where the flow velocity is high. Notably, the angle of rotation does not exceed $5^\circ$. One of these calculation results is shown in Fig.~\ref{figAthree}.

Consider the rotation of a nanotube by the fluid flow when the particle touches the fixing radius with its edge, and let this point be $A_{1}$. In this case, it ceases to move translationally and performs only rotational motion relative to point $A_{1}$. The rotational moment can be approximately determined based on the formula (\ref{eqAone}),
\begin{equation}
\label{eqAsix}
\vec{M} = \int_{A_{1}A_{2}} \left[\overrightarrow{A_{1} A}, \, dm \left.\frac{d\vec{v}}{dt}\right|_{A}\right] \approx \left[\overrightarrow{A_{1} S}, \, \left.\frac{d\vec{v}}{dt}\right|_{S}\right] m,
\end{equation}
where $S$ is some point on the nanotube.
The equation of rotational motion of the nanotube about point $A_{1}$ is
\begin{equation}
\label{eqAseven}
I_{A1} \frac{d \vec{\omega}}{dt}=\vec{M},
\end{equation}
where $I_{A1} = m l^{2} / 3$ is the moment of inertia of the particle about the axis perpendicular to the nanotube and passing through its end $A_{1}$,  $\vec{\omega}$ is the angular velocity of rotation, and $\vec{M}$ is the torque exerted on the particle by the liquid relative to point $A_{1}$. We approximate the torque considering that the nanotube is small and the averaged torque is applied to its center, point $C$:
$$
	I_{A1} \frac{d \vec{\omega}}{dt}\approx m \left[\overrightarrow{A_{1}C}, \,  \left. \frac{d\vec{v}}{dt}\right|_{C} \right].
$$
Integrating over time, assuming the vector $\overrightarrow{A_{1}C}$ is constant over the small time interval $\Delta t$ (numerical calculations of the nanotube rotation angle in $\Delta t = 10^{-4}$~s around the point of contact demonstrate values less than 0.02$^\circ$), we obtain the approximate formula for the $i$-th time step:
$$
	I_{A1} \left. \vec{\omega} \right|_{i} \approx m \left.\left[\overrightarrow{A_{1}C}, \,  \vec{v}(C) \right] \right|_{i},
$$
\begin{multline}
\label{eqAeight}
\frac{m l^2}{3} \left. \omega_{z} \right|_{i} \approx \\ \approx \left. m \left(  \left(x_{C} - x_{A_{1}}   \right)v_{y}(x_{C},\, y_{C})
	- \left(y_{C} - y_{A_{1}}   \right)v_{x}(x_{C},\, y_{C})      \right)
	\right|_{i}.
\end{multline}

If the nanotube has touched another particle at its point $N$, then the rotational moment from the liquid relative to point $N$  is approximately given by [similarly to formula~\eqref{eqAfour}]
\begin{multline}
\label{eqAnine}
\vec{M} =  \int_{A_{1} N} \left[\overrightarrow{N A}, \, dm \left.\frac{d\vec{v}}{dt}\right|_{A}\right] +  \int_{N A_{2}} \left[\overrightarrow{N A}, \, dm \left.\frac{d\vec{v}}{dt}\right|_{A}\right] \approx\\
\approx \left[\overrightarrow{N S_{1}}, \,  \left.\frac{d\vec{v}}{dt}\right|_{S_{1}}\right] \frac{m}{l} \times |NA_{1}| + \left[\overrightarrow{N S_{2}}, \, \left.\frac{d\vec{v}}{dt}\right|_{S_{2}}\right]  \frac{m}{l} \times |NA_{2}|,
\end{multline}
where $S_{1}$ and $S_{2}$ are some points on the nanotube located on either side of point $N$.

The equation of the rotational motion of the nanotube relative to point $N$ is
\begin{equation}
\label{eqAten}
I_{N} \frac{d \vec{\omega}}{dt}=\vec{M},
\end{equation}
where $I_{N} = m l^{2} / 12 + m \left|CN  \right|^{2}$ is the moment of inertia of the particle relative to the axis that is perpendicular to the nanotube and passing through point $N$,
$$
	I_{N} \left. \vec{\omega} \right|_{i} \approx \left( \left. \left[\overrightarrow{N S_{1}}, \,  \left.\vec{v}\right|_{S_{1}}\right]\frac{m}{l}  \times |NA_{1}| + \left[\overrightarrow{N S_{2}}, \, \left.\vec{v}\right|_{S_{2}}\right]  \frac{m}{l} \times |NA_{2}| \right) \right|_{i},
$$
\begin{multline}
\label{eqAeleven}
\left( m\, l^2 / 12 + m \left|CN  \right|^2\right) \left. \omega_z \right|_i \approx \\ \approx \frac{m}{l} \left[\left|N\, A_1\right| \times f_4 (x_{S1},y_{S1}) \right.
	+\left.\left. \left|N \,A_2 \right| \times f_4 (x_{S2},y_{S2}) \right] \right|_{i},
\end{multline}
$$f_4(x,y) = \left(x - x_N \right) \, v_y \left(x,\, y\right)
- \left(y - y_N  \right) \, v_x (x,\, y),$$
$$ \left|N \,A_1 \right| = \sqrt{\left(x_{A1} - x_N \right)^2 + \left(y_{A1} - y_N \right)^2 },$$  $$\left|N \,A_{2} \right| = \sqrt{\left(x_{A2} - x_{N} \right)^{2} + \left(y_{A2} - y_{N} \right)^{2} } ,    $$
$$\left|C\,N  \right| = \sqrt{\left(x_{C} - x_{N} \right)^{2} + \left(y_{C} - y_{N} \right)^{2} }.$$ Formula~\eqref{eqAeight} is obtained from  formula~\eqref{eqAeleven} in the limit when $N\to A_1$.

\section{Coefficient of friction}\label{appendix:frictionCoefficient}
Let us make a rough estimate of the viscous friction coefficient $k_f$ for the nanotubes under consideration. To do this, we need to estimate the Reynolds number, $\mathrm{Re}= d v_c \rho / \eta_0 \approx 10^{-6}$, where the characteristic flow velocity is $v_c \approx$ 10 µm/s \cite{Hamamoto2011}. As the first approximation, we use a linear dependence on nanotube velocity $F_d = k_f v_c$. The drag force is expressed as $F_d = F_{d\mathrm{Lamb}} \left( 1-0.87 \lambda^2 + 0.514 \left( 1 - \exp(-\mathrm{Re})\right) \lambda^3 \right) $, where $\lambda = \left( 0.5 - \gamma - \ln(\mathrm{Re}/8) \right)^{-1}$, $F_{d\mathrm{Lamb}} = 4 \pi \eta_0 v_c \lambda l$, and the Euler constant is equal to $\gamma \approx 0.577$~\cite{Huner1977,Tang2006}. As a result, we obtain $k_f = F_d / v_c \approx 5\times 10^{-10}$ kg/s. When some nanotube rotates in a medium, the medium exerts resistance. Assuming that the drag force is applied to one end of the nanotube, the magnitude of the viscous torque is approximated by the formula $M_\mathrm{vis} = -k_f \, \omega \, (l/2)^2$.

\nocite{*}
\bibliography{KolegovPRE2024}

\end{document}